\documentstyle[preprint,aps,eqsecnum,floats]{revtex}\def\mathcal{\cal}

\input epsf

\begin{document}
\baselineskip=18pt
\bibliographystyle{simpl1} 
\date{\today}
\def\PP{{\cal P}}

\title{Fictitious level dynamics: a novel approach
to spectral statistics in disordered conductors}
 
\author{John T Chalker$^1$, Igor V. Lerner$^2$, and Robert A. Smith$^2$}
\address{$^1$Theoretical Physics, University of Oxford, 1 Keble Road,
Oxford OX1 3NP, United Kingdom\\
$^2$School of Physics and Space Research, University of
Birmingham, Edgbaston, Birmingham~B15~2TT, United Kingdom
}\maketitle \begin{abstract}
 We establish a new approach to calculating spectral statistics in
disordered conductors, by considering how energy levels move
in response to changes in the impurity potential. We
use this fictitious dynamics to calculate the spectral
form factor in two ways. First, describing the
dynamics using a Fokker-Planck equation,
we make a physically motivated  decoupling, obtaining
the spectral correlations in terms of the
quantum return probability. Second, from
an identity which we derive between two- and three-particle
correlation functions, we make a mathematically
controlled decoupling to obtain the same result.
We also calculate weak localization corrections
to this result, and show for two
dimensional systems  (which are of most interest) 
that corrections vanish to three-loop order.
   \end{abstract}\draft

\section{Introduction}
Numerous properties of quantum systems can be described in terms
of their energy spectra. For complex systems  
an exact determination of energy levels is not feasible, and
a statistical description becomes necessary. It turns out that 
the Wigner-Dyson (WD) statistics \cite{Wig,Dsn} 
of eigenvalues of random Hermitian matrices describes energy levels
in a wide variety of different systems
\cite{RMT}.  The joint distribution of eigenvalues is
 dominated by level repulsion
and is universal in the sense that level correlations depend only
upon the symmetry of the Hamiltonian while all specific
properties of the system are absorbed into the mean level spacing, $\Delta$.
A very important feature of WD statistics is that  -- by construction 
of invariant ensembles of random matrices --
spectral properties are independent of eigenstate correlations.
In real systems such an independence can at best be approximate. 
It holds, however, in the ergodic regime where the entire phase 
space of a system is explored. 
If a non-ergodic regime is of interest, not only is WD statistics 
inapplicable but the whole concept of the independence of spectral and
eigenstate correlations should be re-examined. 

Disordered mesoscopic conductors present a natural ensemble for a statistical
description - the ensemble of impurity configurations.
In this case spectral statistics 
in the non-ergodic regime are very important for both transport and 
thermodynamic properties of electrons.  Different regimes in disordered
conductors are determined  by the energy or time  scale,
as shown in Fig.\ 1.
The ergodic regime involves
energy level separations $\varepsilon\alt E_c\equiv
\hbar/t _{\text{erg}}$ where $E_c$ is called the Thouless energy,
and  $ t _{\text{erg}}\sim D/L^2$ is the time
required for the electronic diffusive motion, with diffusion constant $D$,
to fill all phase space, in a sample of size $L$. The quantum limit of this
regime corresponds to smaller energy separations,
  $\varepsilon\alt \Delta$, where
$\Delta$ is the mean level spacing, and to longer
times $t\agt t_{\!_H}$, where $t_{\!_H} \equiv \hbar/\Delta$ 
is the Heisenberg time. 
Note that the ratio $E_c/\Delta$ is proportional to the dimensionless
conductance $g$ (i.e.\ the conductance measured in the units of
$e^2/\hbar$), and is large in the metallic phase.
The diffusive regime involves energies
$\hbar/t _{\text{erg}}\alt \varepsilon\alt \hbar/t_{\text{el}}$,
where $t_{\text{el}}$ is mean elastic-scattering time.
The largest energies, $ \hbar/t_{\text{el}}\alt \varepsilon\alt
\varepsilon_{\!_F}$, (where $\varepsilon_{\!_F}$ is the Fermi energy)
 and the shortest times, $t\alt t_{\text{el}}$,
correspond to the ballistic regime in which multiple scattering by impurities 
is irrelevant. 

\begin{figure}
\epsfxsize=6truein
\epsffile{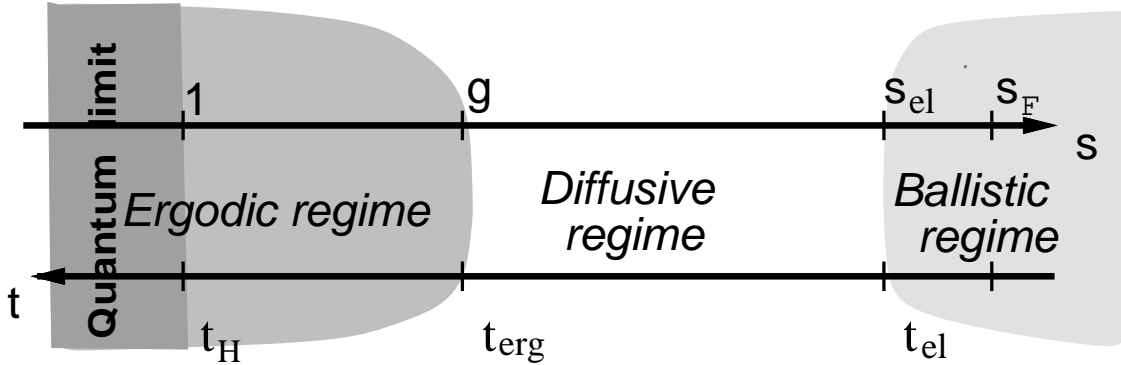}
\vspace{0.5truein}
\caption{Regimes of energy and time in a disordered metal;
here $s_{\text{el}}=\hbar/t_{\text{el}}\Delta $ and
 $s_{\!_F}=\varepsilon_{\!_F}/\Delta$.}
\end{figure}
It was first conjectured by Gor'kov and Eliashberg \cite{GE}
and then shown by Efetov \cite{Ef:83} that spectral 
correlations in the ergodic regime are described by the random matrix
theory (RMT) of Wigner and Dyson. Correlations in the 
non-ergodic diffusive regime
which are important, in particular, for universal conductance 
fluctuations were analysed by Altshuler and Shklovskii \cite{AS}
at leading order in diagrammatic perturbation theory.
Their results were later
reproduced  by Argaman et al \cite{ImSm} within the semiclassical 
approach, using the diagonal approximation. 

We have recently developed \cite{CLS1}
an alternative approach to level statistics
in the non-ergodic  regime  which takes into account the
inevitable coupling between eigenvalue and eigenstate correlations,
and can be extended beyond the region of validity of the
perturbative technique.
Our approach is based on the idea of parametric
motion through the ensemble of disordered Hamiltonians: 
we treat the energy eigenvalues as particles with fictitious
dynamics induced by changing some parameter of the Hamiltonian. 
This dynamics takes the form of Brownian motion in a fictitious time
$\tau$ related to the parameter being changed. Originally, this idea
was employed by Dyson \cite{Dsn2} in the context of RMT. 
Later,  in the context of the semiclassical description, Pechukas
\cite{Pechukas} used motion along a smooth path in the space of Hamiltonians
to generate fictitious dynamics of a different kind.
Both Dyson and Pechukas were interested in level dynamics
primarily as a way of generating the level distribution.
By contrast,
a number of recent authors \cite{para},
notably Szafer, Simons and Altshuler
\cite{ASz,SiA1},\nocite{SiA2} have investigated 
the dynamical problem in its own right,
calculating parametric statistics:
eigenvalue correlations  as a function of position 
in the space of Hamiltonians.
\nocite{Yukawa} In distinction to our approach,
eigenfunction correlations have played no
role in previous work
\cite{para,ASz,SiA1,Yukawa,Bee3},
an assumption justified only for the ergodic regime. 
 
We have found \cite{CLS1}
that a treatment based on Brownian motion through the 
ensemble of Hamiltonians
provides a unified description of all regimes in disordered conductors,
except the quantum limit ($t\agt t_{\!_H}$, or $E\alt \Delta$),
which is also beyond the scope of diagrammatic and semiclassical approaches.
The main result is 
a new relation explicitly linking the spectral correlation 
function to the quantum return probability for an expanding wavepacket
which, in turn, is related to a certain eigenfuction correlator. 
The derivation given in Ref \cite{CLS1}
has a limitation: when obtaining a closed Langevin equation describing
the Brownian motion, we make use of an 
uncontrolled, although physically transparent assumption. Thus, 
within the framework of that calculation
it is not possible to establish exactly the region of validity
and accuracy. 

In this paper, we re-derive the relation between spectral 
and eigenstate correlation functions using a more explicit procedure:
we decouple a certain exact relation between two-
and three-level correlation functions using the Kirkwood superposition 
approximation. Then we examine the accuracy 
of this decoupling using perturbative diagrammatic techniques. 
Remarkably, it not only reproduces the results
of the diagonal approximation \cite{AS,ImSm},
but  holds well beyond it. 
We show this to third order in a perturbative expansion
in $g^{-1}$, for two-dimensional systems with or
without time-reversal symmetry.
We are therefore encouraged to believe
that the connection between spectral
and eigenstate correlation functions
should be useful rather generally, and especially
for problems 
where the usual diagrammatic technique cannot straightforwardly 
be applied, such as spectral statistics 
in the critical regime near the Anderson transition. 

\section{Definitions and main results}

A convenient way to consider spectral correlations is to introduce
fictitious level dynamics in response to changing some parameter
$\lambda$ of the Hamiltonian. 
Thus we parameterize the ensemble of Hamiltonians as follows
\begin{eqnarray}
\label{FPEns}
H(\lambda)&=&H_0+\lambda W(\bbox{r})\,.
\end{eqnarray}
Here both $H_0$ and $H(\lambda)$ belong to the same symmetry class, and
the point $\lambda=0$ corresponds to some arbitrary choice of one of the
many members of the same ensemble. We will specify the choice of 
$H_0$ and $W$ in the next section. 

We consider in this paper the two-level correlation function (TLCF)
and its Fourier transform, the spectral form factor, in a disordered
conductor described by the Hamiltonian Eq.\ (\ref{FPEns}).
Let $E_n(\lambda)$ be the energy levels of
$H(\lambda)$. 
We introduce the density of states per unit volume (DoS) as
\begin{equation}
\label{rho}
\rho(E,\lambda) = {1\over L^d}\sum_n \delta\bigl(E-E_n(\lambda)\bigr)\,.
\end{equation}
The mean level spacing, $\Delta$, is then related to the mean DoS,
$\rho\equiv\langle\rho\rangle$, by $\Delta \!=\! (\rho L^d)^{-1}$.
The TLCF is defined as
\begin{equation}
\label{R(E)}
R(s,\lambda) = 
\rho^{-2} \Bigl< \rho(E+s\Delta,\lambda) \rho(E,0) \Bigr>-1\,,
\end{equation}
 where $\omega=s\Delta$ is the energy difference between two levels.
The mean DoS is practically a constant in the entire energy region
of interest (as it changes only at scale of order $\varepsilon_{\!_F}$ 
while we consider energy windows centered at  $\varepsilon_{\!_F}$ 
of width not exceeding $\hbar/t_{\text{el}}\ll\varepsilon_{\!_F}$). 
We consider only values of $\lambda$ small enough so that the statistical
regime does not change  and neither does the mean DoS
(for large enough $L$ this nevertheless allows
arbitrarily large $\lambda$ on the scale relevant for parametric correlations).
Because of this the TLCF cannot depend on either $E$, 
or on the choice
of the point $H_0$ in the ensemble (\ref{FPEns}). We 
define the (dimensionless) spectral form factor as
\begin{equation}
\label{K(t)}
K(t,\lambda) = \int_{-\infty}^{\infty}
e^{-i s t/t_{\!_H}} R(s,\lambda)\, ds\,.
\end{equation}

Our main result relates the spectral form factor to the quantum 
return probability $p(t)$ of a diffusing electron as follows:
\begin{eqnarray}
\label{K1}
K(t)=
{(2\pi\hbar\rho )^{-1}|t|\,p(t)\over {1+
(\pi\hbar\rho )^{-1} \int_{0^+}^{|t|}\,p(t') d t'}} \,.
\end{eqnarray}
We have obtained this expression for times shorter than the Heisenberg
time $t_{\!_H}\equiv \hbar/\Delta$.
Here we define $p(t)$ as the probability density for the wave packet, 
originally created in a small volume $V_0\sim\ell^{d}$,
to remain in this volume at the time $t$
($\ell$ is  the elastic mean free path
 which is a natural coarse-graining size for the disordered metal;
however, we could choose $V_0$ arbitrarily, provided that 
$\ell^d\agt
V_0\agt\lambda_{\!_F}^d$ where $\lambda_{\!_F}$ is the Fermi wavelength).
The ensemble-averaged return probability
is related, as we will show later, to the following wave-function 
correlations: 
\begin{eqnarray}
\label{p(t)}
p(t)
&=&\int \!\!d^dr\, \left<\sum_l|\psi_n(r)|^2|\psi_{n+l}(r)|^2
e^{-i(E_n-E_{n+l})t/\hbar}\right> \,.
\end{eqnarray}
It is important to note that, by definition of the wave packet above, the
summation here is limited to the number of levels ${\cal N}\sim L^d/V_0$
with energies lying within the energy band of width $E_0\sim 1/\rho V_0$.
  
Equation\ (\ref{K1}) relates the spectral 
and wave-function correlations. Let us analyse this relation in the
metallic phase. 
In the diffusive regime, $t_{\text{el}}\alt t\alt t_{\text{erg}}$,
the quantum return probability  $p(t)$ reduces at leading order to the {\it
classical } return probability for random walks, multiplied by a
symmetry factor $2/\beta$, where  $\beta=1,\,2 \,\text{or}\, 4$ is
the usual index corresponding to the orthogonal, unitary, and symplectic
symmetry ensembles, respectively
 \cite{RMT}:
\begin{eqnarray}
\label{dif}
p_{0}(t)  = {2 \over \beta (4 \pi D t)^{d/2} }\,.
\end{eqnarray}
Noting that in the ballistic regime,  $t\alt t_{\text{el}}$, $p(t)$
saturates at $p_{0}(t_{\text{el}})\sim 1/\ell^{d}$,
one sees that the integral in the denominator of Eq.\ (\ref{K1})
is of order $(t_{\text{el}}\Delta/\hbar) (L/\ell)^d\sim g_0^{-1}$
for $d>2$, and of order $ g_0^{-1}\ln (t/t_{\text{el}})$ for  $d=2$.
It is well known that such an integral describes a weak localization 
correction
to conductance and other physical quantities\cite{LKh2,DK}. On the other hand, 
the quantum return probability  $p(t)$ contains weak localization corrections
itself.
Neglecting  these corrections in both the numerator and denominator of
Eq.\ (\ref{K1}), we reduce it to
\begin{eqnarray}
\label{K3}
K_0(t)&=&
(2\pi\hbar\rho )^{-1}|t|\,p_{0}(t)\,.
\end{eqnarray}
To leading order, this expression is also valid in
 the ergodic regime, $t_{\text{erg}}\alt t\ll t_{\!_H}$, where
the classical return probability saturates at $(2/\beta) L^d$ so that
the second term in the denominator in
Eq.\ (\ref{K1}) is of order $t/t_{\!_H}\ll1$  and  may be neglected.
We should not expect Eq.\ (\ref{K1}) to be correct in the quantum limit,
$t\gg t_{\!_H}$, as we have derived it under the assumption that the opposite inequality
holds, as will be seen later. Indeed, in this regime Eq.\ (\ref{K1}) 
gives the saturation of $K(t)$ at $1/2$ instead of the correct limiting
value $K(t)=1$.

Equation (\ref{K3})
 coincides with the result obtained by Argaman et al \cite{ImSm},
using the diagonal approximation in semiclassical periodic-orbit
theory.
The Fourier transform of this expression corresponds to
 the TLCF obtained originally by Altshuler and Shklovskii\cite{AS}.
In the diffusive regime, $R(s,0)
\sim A_d \,g^{-d/2} s^{d/2-2}$, where
$A_d$ is a numerical coefficient which is zero
for $d\!=\!2$\cite{KL:94}, and in the ergodic regime,
$R(s,0) \sim -1/s^2$ \cite{bmcom2}.

The second relation we obtain between $K(t)$ and $p(t)$ is:
\begin{eqnarray}
\label{K4}
K(t)+(\pi\hbar\rho)^{-1}\int_{0+}^{|t|}K(t-t')p(t')dt'=
(2\pi\hbar\rho)^{-1}|t|p(t)
\end{eqnarray}
which we can see is very similar to Eq.\ (\ref{K1}). The latter is
obtained from a diagrammatic analysis discussed in section (VI). The
new feature is that we have a convolution of $K(t)$ and $p(t)$ which
occurs because the decoupling is in $\omega$-space rather than $t$-space.
In 2d, up to 3-loop
order in perturbation theory, both these relations correctly
reproduce the TLCF.
It seems to us quite remarkable that a relation derived from a
phenomenological model of energy level dynamics could be exact to 3-loop
order. We note that the 2d
case is expected to be a good model for the behaviour of the system for
$d>2$ at the mobility edge. From the point of view of a power-counting
analysis of the properties of $K(t)$ at the mobility edge, both
Eq.\ (\ref{K1}) and Eq.\ (\ref{K4}) should work equally well.

\section{Random Walks through the Ensemble of Hamiltonians}
 We have established elsewhere \cite{CLS1} the relation (\ref{K1}),
using a Langevin equation to describe the motion of levels on the
energy axis in response to a random walk through the
 the ensemble of Hamiltonians.
In contrast to eigenvalue statistics in  RMT
where the Brownian motion ideas were originally applied \cite{Dsn2,Bee3},
the Langevin equation for level motion is not closed,
and certain assumptions are
required to solve it.
In the following section, we will re-derive Eq.\ 
(\ref{K1}) with the help of a different approach based on the decoupling
of a certain exact relation between two- and three- level correlation 
functions. Before doing this, however, it is useful to analyze the Brownian
motion picture within the Fokker-Planck scheme. Although the 
Langevin and Fokker-Planck schemes are in principle equivalent,
the assumptions required in order to make the description closed
 are different.
The Langevin scheme of Ref.\ \onlinecite{CLS1} is physically more 
transparent. The advantage of the  Fokker-Planck scheme which we 
develop here is that
the approximation made are more closely related
to those analysed in subsequent sections.

We consider paths of two types
 through the ensemble of Hamiltonians (\ref{FPEns})
which, for free electrons in a random potential, have the form 
\begin{eqnarray}
\label{AndHam}
H(\lambda) &=& -{\hbar^2\over 2m}\nabla^2 +U(\bbox{r})+\lambda W(\bbox{r})\,.
\end{eqnarray}
Here both $U(r)$ and $W(r)$ are chosen to be of Gaussian white-noise
form with zero average and
\begin{eqnarray}
\label{UU}
\begin{array}{rclrcl}
\bigl< U(\bbox{r})U(\bbox{r}')\bigr>&\!
=\!&\displaystyle{\hbar\over 2\pi\rho t_{\text{el}}}
 \delta(\bbox{r}\!-\!\bbox{r}')\,,\\[6pt]
\bigl< W(\bbox{r})W(\bbox{r}')\bigr>&
\!=\!&v^2L^d\,\delta(\bbox{r}\!-\!\bbox{r}' )\,.
\end{array}
\end{eqnarray}
The first type of path is a straight line
through the ensemble, and generated by varying $\lambda$ in Eq.\ \ref{AndHam}.
 The second type is a Brownian path through
the ensemble, parameterized by the
fictitious time $\tau$, generated in the following way.
\begin{eqnarray}
\label{LangEns}
H(\tau)&=&H_0+\int^{\tau}_0 \!\!\! d\tau'\, V(\tau',\bbox{r})\,.
\end{eqnarray}
We take $V(\tau,\bbox{r})$ to be
Gaussian distributed with zero average and
\begin{eqnarray}
\label{VV}
\bigl< V(\tau ,\bbox{r})V(\tau' ,\bbox{r}')\bigr>&
\!\!=\!\!&v^2L^d\,\delta(\tau-\tau')\,\delta(\bbox{r}\!-\!\bbox{r}' )\,.
\end{eqnarray}
Referring to Fig 2, one sees that the two ways of exploring the ensemble
are equivalent if one makes the identification $\tau = \lambda^2$.
\begin{figure}
\epsfxsize=0.8\textwidth
\hspace*{\fill}\epsffile{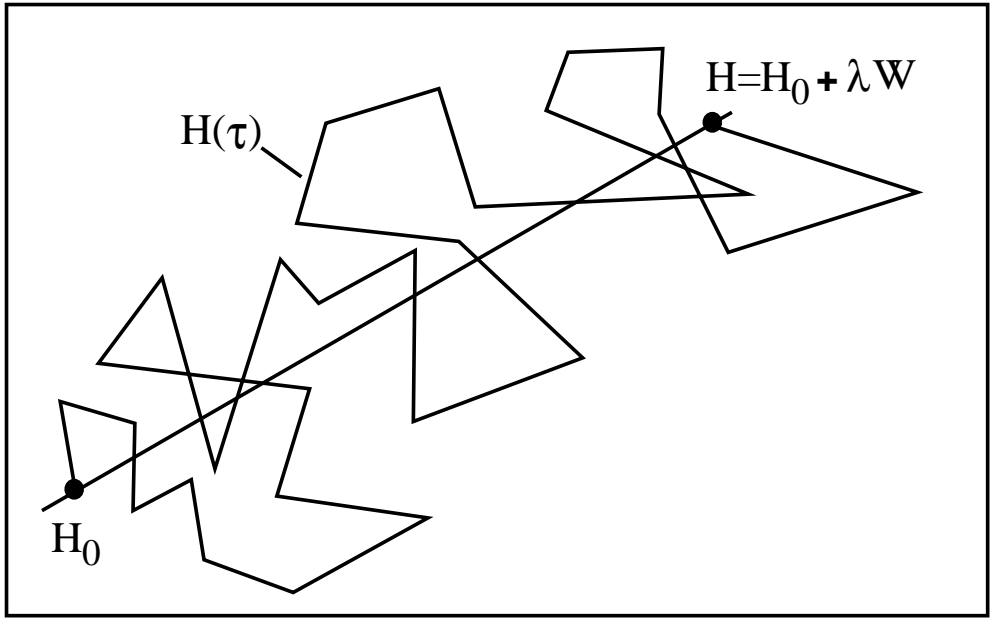}\hspace*{\fill}
\vspace{0.4truein}
\caption{Smooth [$H(\lambda)$] and Brownian [$H(\tau)$] paths through
the space of Hamiltonians}
\end{figure}

In our derivation we will use averages over both $W$
 or, equivalently, $V$ and  then over $H_0$, and we
must here discuss the role of each. The average over all possible
perturbations, $W$, will be necessary to derive the equation of motion
for the of the energy levels, 
and thence the density of states. We
can then obtain correlation functions for energy levels of the system
at different parameter values by averaging over the starting point $H_0$.
Such functions should then depend only upon energy and parameter
differences by homogeneity.

The first step in our derivation is to obtain the equation of motion
for the joint probability density function (JPDF) 
 of energy levels, $P\bigl(\{E_n\},\tau\bigr)$.
 We use perturbation theory to second order to
calculate the change of $E_n(\tau)$ in response to the evolution from
$\tau$ to $\tau+\delta\tau$. After averaging over $W$ we obtain
\begin{mathletters}
\label{ECorr}
\begin{eqnarray}
\label{delE}
\Bigl<\delta E_n(\tau)\Bigr> = (\delta\tau)v^2\sum_{m\ne n}
\frac{c_{nm}(\tau)}{E_n(\tau)-E_m(\tau)}\\
\label{delEE}
\Bigl<\delta E_n(\tau)\delta E_m(\tau)\Bigr> = v^2 c_{nm}(\tau)
\end{eqnarray}
\end{mathletters}
where
\begin{eqnarray}
\label{c}
c_{nm}(\tau)=L^d\int \!d^dr\, \bigl|\psi_n(\tau,r)\bigr|^2 \bigl|
\psi_m(\tau,\bbox{r})\bigr|^2,
\end{eqnarray}
and $\psi_n(\tau,\bbox{r})$ are the corresponding 
eigenfunctions of $H(\tau)$.
Before we can use the above equations to derive a Fokker-Planck equation
for the JPDF, $P\bigl(\{E_n\},\tau\bigr)$, we must make an assumption.
We replace $c_{nm}(\tau)$ by its average over the ensemble of $H_0$,
This amounts to ignoring correlations between eigenvalues and 
eigenvectors. We take the disorder average to be a function only of the
energy difference, $\omega=E_n-E_m$, within the window of interest:
\begin{equation}
\label{c(E)}
\bigl< c_{nm}(\tau) \bigr> \equiv c(\omega )\,.
\end{equation}
Furthermore, we assume that in the Fourier transform of the wavefuntion
correlator $c(\omega)$ we may neglect the correlations between the 
eigenvectors and eigenvalues so that 
\begin{eqnarray}
\label{C(t)}
C(t)=\left<\frac{\Delta}{2\pi\hbar}
\sum_l c_le^{-i\Delta l t/ \hbar}\right>=
\int_{-\infty}^{\infty} c(\omega)e^{-i\omega t/\hbar}
\frac{d\omega}{2\pi\hbar}\,.
\end{eqnarray}
to the return probability of a diffusing electron, Eq.\ (\ref{p(t)}).
To this end, consider a wavepacket made from the eigenstates of
$H(\tau)$ and concentrated initially in a volume $V_0$ of order $\ell^d$
 near the origin
(since $\tau$ plays no role, we suppress it as a label in the following):
\begin{eqnarray*}
\Psi({\bbox{r}},t) =A \sum_n \psi_n({\bbox{0}})^*
\psi_n({\bbox{r}}) e^{-iE_n t / \hbar}\,.
\end{eqnarray*}
Here the summation is limited to ${\mathcal N}\sim (L/\ell)^d$
 levels with energies
$|E_n|\alt  1/\rho \ell^d$,
and the normalization constant is $A^2=L^d/{\mathcal N}$.
The ensemble-averaged return probability
$p(t)  =\left<|\Psi({\bbox{0}},t)|^2\right>$  is given by
\begin{eqnarray*}
p(t)&=&A^2\sum_{nm}\left<|\psi_n(0)|^2|\psi_m(0)|^2
e^{-i(E_n-E_m)t/\hbar}\right>=\left<\sum_l|\psi_n(0)|^2|\psi_{n+l}(0)|^2
e^{-i(E_n-E_{n+l})t/\hbar}\right> \,,
\end{eqnarray*}
where we have used the fact that the first sum above
depends only on the difference $|n-m|$. Noticing also that the
ensemble-averaged quantity is spatially homogeneous, we reduce this expression
to that given in Eq.\  (\ref{p(t)}). Comparing this to the
definition of $c_{nm}$,  Eq.\    (\ref{c}), we obtain for $t>0$
\begin{eqnarray}
p(t)  &=&\frac{1}{L^d}\left<\sum_l c_{n,n+l}e^{-i(E_n-E_{n+l})t/\hbar}
\right> \,.
\label{p=C}
\end{eqnarray}
On the face of it, this coincides,  up to a constant factor, with  $C(t)$,
the Fourier transform of $c_{n,n+l}$, introduced in Eq.\ (\ref{C(t)}).
There is, however, an essential difference: $C(t)$ is defined by
the Fourier sum containing
all the levels (say, up to $\varepsilon_{\!_F}$), while
the Fourier sum for $p(t)$ contains only the levels within the bandwidth $E_0
\ll\varepsilon_{\!_F}$. When $|E_n-E_{n+l}|\agt E_0$ the two levels are
practically
uncorrelated, and $c(\omega)=1$ for $\omega\agt E_0$, so that $C(t)$ contains
a
$\delta$-like function for $t$ near zero. As we are not interested in
an exact description at the ballistic  time scale, we can represent the
relation between  $C(t)$ and $p(t)$ as follows:    %
\begin{eqnarray}
\label{pC}
\begin{array}{rcl}
p(t)&=&2\pi\hbar\rho\,C(t)\,,\qquad t>0\,;\\
\displaystyle{\int_0^t C(t')dt'}&=&\displaystyle
{1\over2}+{1\over2\pi\hbar\rho}\int_{0^+}^t p(t')dt' \,.
\end{array}
\end{eqnarray}

We also note that the definition of $H(\tau)$ in Eq.\ (\ref{LangEns})
causes the energy levels to move away from each other indefinitely as
parametric time increases. To overcome this problem we introduce a
rescaling term, $-\delta\tau \,U(E_n)$ to the r.h.s. of Eq.\ (\ref{delE}).
This $U(E_n)$ can be thought of as a Lagrange multiplier, and it will be
set later on by the condition that correlation functions can depend only
on differences in parametric time. 

With these considerations, starting with Eqs.\ (\ref{ECorr}) we end
up with the Fokker-Planck equation for JPDF:
\begin{eqnarray}
\label{FPeqn}
\frac{1}{v^2}
\frac{\partial P}{\partial\tau}=-\sum_n\frac{\partial}{\partial E_n}
\left(\frac{\partial{\mathcal U}}{\partial E_n}P\right)
+\sum_{nm}\frac{\partial^2}{\partial E_n\partial E_m}
\left(c_{nm}P\right)
\end{eqnarray}
where the drift potential term, ${\mathcal U}$ is the sum of one-particle
and two-particle potentials,
\begin{eqnarray}
\label{U}
{\mathcal{U}}(\{E_n\})=\sum_n U(E_n)+\frac{1}{2}\sum_{m\ne n}f(E_n-E_m).
\end{eqnarray}
The one-particle potential arises from the energy rescaling described
above, and may be considered as a confinement potential for a 
one-dimensional gas of fictitious particles interacting via the 
two-particle potential, $f(\omega)$, which is related to $c(\omega)$ by
\begin{eqnarray}
\label{fc}
\frac{\partial f(\omega)}{\partial\omega}=
\frac{c(\omega)}{\omega}.
\end{eqnarray}
We see that both the drift potential and the diffusion term in the 
Fokker-Planck equation are expressed in terms of the function 
$c(\omega)$, which we have shown to be related to the return probability
$p(t)$. The off-diagonal diffusion terms in Eq.\ (\ref{FPeqn}), which are
due to eigenfunction correlations, mean that the static solution does
not have a simple Gibbs form. 
In fact we cannot write down its solution in closed form at all.
The absence of a simple static 
solution to the Fokker-Planck equation (\ref{FPeqn})
is an important difference between the current problem
and the Brownian motion approach to RMT \cite{Dsn2}
  where such a solution yields  the exact JPDF. 
However, the JPDF contains much more
information than we require; for our purposes it is sufficient
to study the equation of motion for the density of states,
which can be written in the form
\begin{eqnarray}
\label{DoS}
\rho(E,\tau)=L^{-d}\overline{\sum_n\delta(E-E_n)^{\raisebox{-10pt}{$\,$}}},
\end{eqnarray}
where $\overline{^{^{\,}}\ldots^{^{\,}}}$ means averaging over the
JPDF $P\bigl(\{E_n\},\tau\bigr)$.
Following the procedure of Dyson\cite{Dsn3} and Pastur\cite{Pastur} we obtain, 
\begin{eqnarray}
\label{Dys}
\frac{1}{v^2}
\frac{\partial\rho(E,\tau)}{\partial\tau}=
\frac{\partial^2}{\partial E}[c(0)\rho(E,\tau)+
\frac{\partial}{\partial E}\left[\frac{\partial U}{\partial E} +
L^d \int dE'\rho_2(E,E',\tau)\frac{\partial f(|E-E'|)}{\partial E'}
\right],
\end{eqnarray}
where
\begin{eqnarray}
\label{rho2}
\rho_2(E,E',\tau)
=L^{-2d}\overline{\sum_{n\ne m}\delta(E-E_n)\delta(E-E_m)^{\mbox{$\,$}}}
\end{eqnarray}
Since we are only interested in level correlations in energy windows
small compared to the scale at which $\rho$ varies, the first term in
Eq.\ (\ref{Dys}) is negligible. We rewrite the last term in 
Eq.\ (\ref{Dys}) in terms of the static TLCF, $R(E-E')$,
\begin{eqnarray}
\label{DoS2}
\rho_2(E,E',\tau)=\rho(E,\tau)\rho(E',\tau)[1+R(E-E')].
\end{eqnarray}
After substitution into Eq.\ (\ref{Dys}) we see that the integral over
$\rho\rho R$ is dominated by a region of relatively small energy 
differences since the product $R f'$ falls off rapidly. In this region
$\rho$ is roughly constant, and the integral vanishes by oddness of the
integrand. We have therefore arrived at the equation
\begin{eqnarray}
\label{DysPa}
\frac{1}{v^2}
\frac{\partial\rho(E,\tau)}{\partial\tau}=
\frac{\partial}{\partial E}\left[\rho(E,\tau)\frac{\partial}{\partial E}
\left(U(E)+L^d \int dE' \rho(E',\tau)f(|E-E'|)\right)\right]
\end{eqnarray}
This is a non-linear equation, and to proceed further we must linearize 
it. The static solution of Eq.\ (\ref{DysPa}) is just the equilibrium
density of states, $\rho_{eq}(E)$, and we perform our linearization
around by expanding around $\rho_{eq}(E)$, to give the following
equation for $\widetilde\rho(E,\tau)=\rho(E,\tau)-\rho_{eq}(E)$:
\begin{eqnarray*}
\frac{1}{v^2}
\frac{\partial\widetilde\rho(E,\tau)}{\partial\tau}=
{1\over\Delta}\frac{\partial}{\partial E}\int dE' f(|E-E'|)
\frac{\partial}{\partial E'}\widetilde\rho(E',\tau),
\end{eqnarray*}
where we have approximated $\rho_{eq}(E)$ by $\rho$. Multiplying both
sides of this equation by $\rho(E'',0)$ and averaging over the starting
point $H_0$ gives us the evolution equation for the TLCF, $R(E,\tau)$,
\begin{eqnarray}
\label{Rdyn}
\frac{\partial}{\partial\tau}R(\omega,\tau)=
\frac{\partial}{\partial \omega}\int {d\omega'\over2\pi\hbar}
  f(|\omega-\omega'|)
\frac{\partial}{\partial \omega'} R(\omega',\tau)\,.
\end{eqnarray}
where we have fixed the units of $\tau $ by setting $v^2={\Delta
/\pi\hbar}$.
Eq.\ (\ref{Rdyn})
 can then be solved by taking the Fourier transform to yield the
result for the spectral form factor
\begin{eqnarray}
\label{SFF}
K(t,\tau)=K(t,0)\exp \!\left[{-{M(t) \over 2 \hbar^2  }\,|t\,\tau|}\right]\,,
\end{eqnarray}
where $M(t)=2t f(t)$. From the definition of $f(\omega)$,
Eq.\ (\ref{fc}), we see that $M(t)$ is related to $C(t)$, the Fourier
transform of $c(\omega)$, by
\begin{eqnarray}
\label{FC}
M(t)=2\int_0^t C(t')dt'=
1 \!+\! {1 \over\pi\hbar\rho } \int_{0^+}^{|t|}\!\! p(t') d t\,.
\end{eqnarray}

Although Eq.\ (\ref{SFF}) gives the parametric dependence of $K(t,\tau)$
in terms of a function $M(t)$ related to eigenfunction correlations, we
still do not know $K(t,0)$. To relate $K(t,0)$ to eigenfunction
correlations we introduce a Ward identity
as follows.
Similarly to the TLCF, Eq.\ (\ref{R(E)}), we define
the current-current correlation function:
\begin{eqnarray}
\label{jj}
{\cal C}(s,\lambda)=\Delta^2\sum_{n,m}
\left<\dot E_n(\lambda)\,\dot E_m(0)\,
\delta(E+s \Delta-E_n(\lambda))\,\delta(E-E_m(0))\right>
\end{eqnarray}
The assumption that both the 
correlation functions depend only upon energy and parameter differences 
leads to the  Ward identity
\begin{eqnarray}
\label{ward}
\frac{\partial^2 {\cal C}(\omega,\lambda)}{\partial \omega^2}=
\frac{\partial^2 R(\omega,\lambda)}{\partial \lambda^2}.
\end{eqnarray}
and thence to the relation
\begin{eqnarray}
\label{ward1}
\frac{\partial K(t,\tau)}{\partial\tau}=
-\frac{\Delta}{2}\int\! d\omega\, {\cal C}(\omega,0)e^{-i\omega t/\hbar}.
\end{eqnarray}
Finally we can relate ${\cal C}(\omega,0)$ to eigenstate correlations
assuming, as above,  that we can ignore higher order correlations 
between eigenvalues and eigenstates:
\begin{eqnarray}
\label{c=c}
 {\cal C}(\omega,0)=\Delta^2\sum_{n\ne m}\Bigl<W_{nn}\,W_{mm}\,
\delta(\omega-E_n)\,\delta(E_m)\Bigr> \approx v^2 c(\omega)
\end{eqnarray}
from which it follows that
\begin{eqnarray}
\label{K(t,0)}
K(t,0)=\frac{1}{2}\left(\frac{t}{\hbar}\right)^2
\frac{C(t)}{M(t)}.
\end{eqnarray}
With allowance for Eqs.\ (\ref{pC}) and \ref{FC}), this is equivalent to the
relation  (\ref{K1}) obtained within the Langevin picture
of Ref.\ \cite{CLS1}.
 
The crucial assumptions
used in the Brownian motion approach were in the linearization of appropriate
equations, and in the neglection of higher order correlations
between eigenvalues and eigenstates.
We believe that these assumptions 
are reasonable provided that one considers
only behaviour at energy scales much larger than the mean level spacing.
However, it is not possible to establish exactly their region of validity
and accuracy within the Brownian-motion approach developed
here and in Ref.\ \onlinecite{CLS1}. This we will do in the 
next section using an alternative approach.

\section{Exact relations between the return probability and higher-order
correlation functions}

Our aim now is to re-derive Eq.\ (\ref{K1}) using some exact relations
which involve the return probability by making only {\it explicit} 
assumptions
whose region of validity can later be verified.  We again consider the
ensemble of Hamiltonians, each describing a particular realization of the 
impurity potential. Now it will be more convenient to use the 
representation of Eq.\ (\ref{FPEns}) where the path through the 
ensemble is a straight line. We define the Fourier
transform of DoS as follows:
\begin{eqnarray}
\label{E-lam}
{\cal {R}}(t,\lambda)= { L^d} \int_{-\infty}^{\infty}
\rho(E,\lambda) e^{-iEt/\hbar} \, dE=
\sum_n e^{-i E_n(\lambda)t/\hbar}\,.
\end{eqnarray}
As all members of the ensemble are statistically equivalent,
averaging over realizations should give the same result whatever point
along the path has been chosen. Thus one should have
\begin{eqnarray}
\label{lambda}
\Bigl<{\cal {R}}(t,\lambda) {\cal {R}}(t',\lambda)\Bigr>
=
\Bigl<{\cal {R}}(t,0) {\cal {R}}(t',0)\Bigr>
\end{eqnarray}
Now we write for small $\lambda$
$$
{\cal {R}}(t,\lambda) = {\cal {R}}(t,0) + \lambda\dot {\cal {R}}(t,0)+
\frac{1}{2} \lambda^2 \ddot {\cal {R}}(t,0) + O( \lambda^3)
$$
and substitute this expansion into Eq.\ (\ref{lambda}) to obtain
\begin{eqnarray}
2\left<\dot {\cal {R}}(t,0) \dot {\cal {R}}(t',0) \right>
+ \left<\ddot {\cal {R}}(t,0) {\cal {R}}(t',0) \right> +
 \left<\ddot {\cal {R}}(t',0) {\cal {R}}(t,0) \right>=0\,.
\label{dot}
\end{eqnarray}
We will use this identity to derive exact relations between spectral
correlation functions.
It follows from the definition (\ref{E-lam}) that
\begin{eqnarray}
\begin{array}{lcl}
\dot {\cal {R}}(t,0)&=&\displaystyle 
-{it\over\hbar}\sum_n \dot E_n e^{-i E_nt/\hbar}\,,\\[12pt]
\ddot {\cal {R}}(t,0)&=&\displaystyle 
-{it\over\hbar}\sum_n
\left(\ddot  E_n - {it\over \hbar} \dot E_n^2\right) e^{-i E_nt/\hbar}\,,
\label{calEdot}
\end{array}
\end{eqnarray}
where
\begin{eqnarray*}
\dot E_n = \langle\,n|W|n\,\rangle\,,\qquad
\ddot E_n 
= {\sum_m}'\,{\bigl|\,\langle\,n|W|m\,\rangle\,
\bigr|^2 \over E_n - E_m}\,.
\end{eqnarray*}
First consider $\bigl<\dot {\cal {R}}(t)\dot {\cal {R}}(t')\bigr>$.
Averaging over $W$ only, we have
$\left<\dot E_n\,\dot E_{n+l}\right>_{\!_W}=v^2L^{-d} c_{n,n+l}$. 
Hence, averaging
also over $H_0$, we obtain
\begin{eqnarray}
\left<\dot {\cal {R}}(t) \dot {\cal {R}}(t')\right>=
{2\pi v^2 t^2\over\hbar \Delta } \delta(t\!+\!t')\left<{1\over L^d}
\sum_l c_{n,n+l} e^{-i(E_n-E_{n+l})t/\hbar}\right>
={2\pi v^2 t^2\over\hbar \Delta } \, p(t)\,,
\label{dotdot}
\end{eqnarray}
where we have used Eqs.\ (\ref{p=C}) and (\ref{pC}) to relate the average
above to the return probability $p(t)$. 
The crucial assumption in the derivation of Eq.\ (\ref{dotdot})  was that
of homogeneity in energy space which means that 
$\langle c_{n,n+l}\rangle$ does not depend on $n$. We expect this
assumption to be valid in the whole energy range of interest since the
mean density of states is a disorder-independent constant in the energy
window centered  near $\varepsilon_{\!_F}$ of width $E_0\ll\varepsilon_{\!_F}$.

Next, consider $\bigl<\ddot {\cal {R}}(t){\cal {R}}(t')\bigr>$,
initially averaging only over $W$. Noting that 
 $\Bigl<\ddot {\cal {R}}(t){\cal {R}}(t')\Bigr>_{\!_W}={\cal {R}}(t')
\Bigl<\ddot {\cal {R}}(t)\Bigr>_{\!_W}$, as ${\cal {R}}(t')\equiv
{\cal {R}}(t', 0)$ does not depend on $W$, and
\begin{eqnarray*}
\left<\ddot E_n\right>_{\!_W}&=&\displaystyle
\frac{2v^2}{L^d}\sum_{l\ne0}\,{c_{n,n+l} \over E_n - E_{n+l}}\,,
\end{eqnarray*}
we obtain from Eq.\ (\ref{calEdot})
 after rearranging the terms in the summation
\begin{eqnarray*}
\left<\ddot {\cal {R}}(t)\right>_{\!_W}&=&
-{it v^2 \over  \hbar L^d} \sum_n \left\{
\sum_{l\ne0}\,{1-e^{i(E_n-E_{n+l})t/\hbar}\over E_n - E_{n+l}}c_{n,n+l}
-{i t\over\hbar} c_{nn}\right\}e^{-iE_n t/\hbar}\\&=&
-{tv^2 \over\hbar^2 L^d} \sum_{nl}e^{-iE_n t/\hbar}
\int_0^{t } dt'' e^{i(E_n-E_{n+l})t''/\hbar}{c_{n,n+l}}\,.
\end{eqnarray*}
Hence, averaging on $H_0$ as well and using the same assumption as in 
the derivation of Eq.\ (\ref{dotdot}), we have
\begin{eqnarray}
 \left<\ddot {\cal {R}}(t,0) {\cal {R}}(t',0) \right>=
-{2 \pi v^2 t\over \hbar \Delta } \delta(t\!+\!t')
\int_0^{t } \!\!dt''  \left<{1\over L^d}\sum_{lm} c_{n,n+l}\,
e^{i(E_n-E_{n+l})t''/\hbar} \,e^{-i(E_n-E_{n+m})t/\hbar} \right>
\label{0ddot}
\end{eqnarray}
On exchanging $t$ and $t'$, one gets the same expression. 

Now, substituting Eqs.\ (\ref{dotdot}) and (\ref{0ddot}) into Eq.\ (\ref{dot}),
we obtain the following exact relation,
\begin{eqnarray}
\label{p=Q}
tp(t) = Q(t)
\equiv 
\int_0^{t} dt'  \left< {1\over L^d }\sum_{lm} c_{n,n+l}\,
e^{i(E_n-E_{n+l})t'/\hbar} \,e^{-i(E_n-E_{n+m})t/\hbar} \right>
\end{eqnarray}
where $p(t)$ is the return probability, defined by Eq.\ (\ref{p(t)}).
This exact relation is a starting point to investigate
 the nature of the approximations needed to
obtain our Brownian motion formula, Eq.\ (\ref{K1}). The exact relation
involves the higher order correlations, and we need to construct
some decoupling procedure. This will be done in the next section. 
Then we will use the diagrammatic technique to analyze the accuracy of 
the decoupling.

\section{Decoupling of the higher-order correlations
in \lowercase{$\omega$}-Space and \lowercase{$t$}-Space}

 It is seen from Eq.\ (\ref{p(t)}) that $p(t)$ involves
correlations of two energy levels, whereas, from Eq.\ (\ref{p=Q}),
$Q(t)$ involves correlations of three energy levels. We should therefore
assume that $Q(t)$ can be factorized in terms of two-level correlations.
In order to see how to factorize Eq.\ (\ref{p=Q}),
let us note that, using  homogeneity of the energy space and the definitions
of the  DoS and TLCF, Eqs.\ (\ref{rho}) and (\ref{R(E)}), one can represent
the TLCF (at $\lambda=0$) as%
 \begin{eqnarray*}
   R(\omega)\equiv R(s\Delta)=
\Delta\,\left<\sum_m\delta( \omega-E_n+E_{n+m})\right>-1\, .
 \end{eqnarray*}
Then, from the definition  (\ref{K(t)})
one obtains the following representation for the form factor:%
 \begin{eqnarray}
 \label{FF}
 K(t)=\left<\sum_m e^{-i(E_n-E_{n+m})t/\hbar}\right>-{2\pi\hbar\over
\Delta}\delta(t)\,.
 \end{eqnarray}
The $\delta$ function above cancels that arising from summation over high
levels in Eq.\ (\ref{FF}). Naturally, $K(t)$ defined by Eq.\ (\ref{K(t)})
as the form factor for the irreducible TLCF,  Eq.\  (\ref{R(E)}), is 
regular at $t=0$.   
It is seen now that the natural factorization of Eq.\ (\ref{p=Q}) is 
\begin{eqnarray}
\label{Qfac}
Q(t) \approx {2}\int_0^t dt' \left< {1\over L^d }\sum_l c_{n,n+l}
e^{i(E_n-E_{n+l})t'/\hbar} \right> 
\left<\sum_m e^{-i(E_n-E_{n+m})t/\hbar}\right>\,.
\end{eqnarray}
The only feature that deserves comment is the factor of $2$ on the r.h.s.\
of Eq.\ (\ref{Qfac}). 
To see how this arises we rewrite Eq.\ (\ref{p=Q}) using the definition
of $c_{n,n+l}$,
\begin{eqnarray*}
Q(t) = \int_0^t dt'\int \!\!d^dr\,  \left< \sum_{l,m}
|\psi_n(\bbox{r})|^2\,|\psi_{n+l}(\bbox{r})|^2 \,e^{i(E_n-E_{n+l})t'/\hbar}
\,e^{i(E_{n+m}-E_n)t/\hbar} \right>
\end{eqnarray*}
This can be represented on the energy axis in the schematic form shown
in Fig. 3.
\begin{figure}
\epsfxsize=6truein
\epsffile{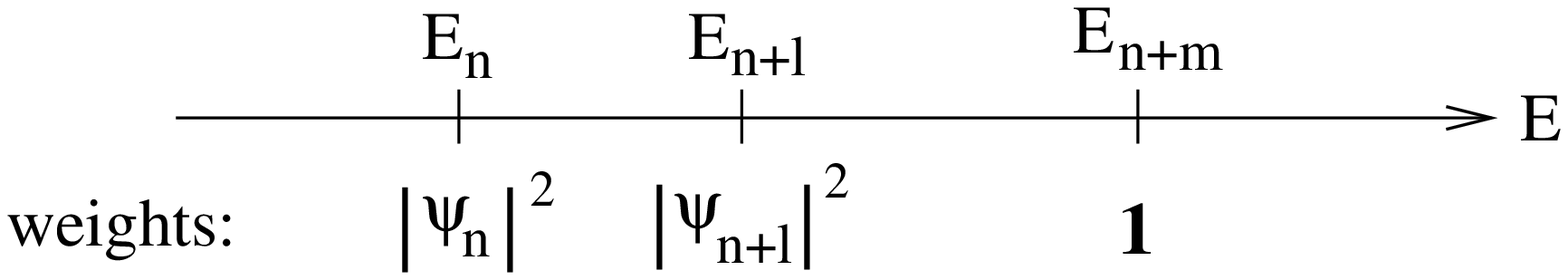}
\vspace{0.5truein}
\caption{Structure of the three level correlation function $Q(t)$.
We see that if in factorization we consider the correlation between
$E_n$ and $E_{n+m}$, we must also consider that between $E_{n+l}$
and $E_{n+m}$.}
\end{figure}
We should allow both for correlations of $E_{n+m}$ with $E_n$
and $E_{n+m}$ with $E_{n+l}$, since these are equivalent. This is made
manifest by changing variable in the integral from $t'$ to $t''=t-t'$,
\begin{eqnarray*}
Q(t) = \int_0^t dt''\int \!\!d^dr\,  \left< \sum_{l,m}
|\psi_n(\bbox{r})|^2\,|\psi_{n+l}(\bbox{r})|^2\, e^{i(E_{n+l}-E_n)t''/\hbar}
\,e^{i(E_{n+m}-E_{n+l})t/\hbar} \right>
\end{eqnarray*}
We rewrite  Eq.\ (\ref{Qfac}), taking into account that (i) $Q(t) = tp(t)$,
(ii) the first average in Eq.\ (\ref{Qfac}) equals $2\pi\hbar \rho\,C(t')$,
  Eq.\ (\ref{C(t)}), and 
 (iii) the second average is equal to $K(t)$, 
Eq.\ (\ref{FF}):
\begin{eqnarray}
\label{Kfac} 
\displaystyle
K(t)=\frac{1}{4\pi\hbar \rho}\,
\frac{ t p(t)}
 { \int_0^t dt'\,
C(t')}\,
\end{eqnarray}
which is, with allowance for Eq.\ (\ref{pC}),
 exactly equivalent to the Brownian motion result of Eq.\ (\ref{K1}).
 Now it is clear that the assumptions we have 
made to derive  Eq.\ (\ref{K1}) 
are equivalent to neglecting three-level
correlations and keeping  only two-level correlations. In the above 
derivation, we have disregarded the $\delta$ function coming from 
Eq.\ (\ref{FF}), as the exact relation  (\ref{p=Q}) has been actually
derived from Eqs.\ (\ref{dotdot}) and (\ref{0ddot}) as $t^2 p(t) = t Q(t)$,
and this  $\delta$ function enters in combination $t \delta(t)$.

For further analysis of the accuracy of Eq.\ (\ref{K1}), it 
will be useful to see how the factorization (\ref{Qfac}) arises
in the energy representation.
We begin by representing $Q(t)$ in Eq.\ (\ref{p=Q}) as follows:
 \begin{mathletters}
\label{Q}
\begin{eqnarray}  \label{QFT}
Q(t)& =& \displaystyle\int_0^t \!\!dt'
\int \! d\omega \int  \!d\omega ' \,  e^{-i\omega 't'/\hbar}\,
e^{i\omega t/\hbar} \,Q(\omega ',\omega )  \,,
 \\
\label{QEE}
Q(\omega ',\omega )&=& 
L^{-d}\left<\sum_{l,m}
\delta(E_{n+l}\!-\!E_n\!-\! \omega ')\,
\delta(E_{n+m}\!-\!E_n\!-\! \omega )\,c_{n,n+l}\right>\,.
\end{eqnarray}
  \end{mathletters}
The function $Q(\omega ',\omega )$ can now be related to the three-level 
correlation function 
${\cal Q}(E'',E',E)$ defined by
  \begin{mathletters}
\begin{eqnarray}
\label{Q1}
 {\cal Q}(E'',E',E)&=&\displaystyle
L^{-d}\left<\sum_{n,l,m}\delta(E''\!-\!E_n)\,
\delta(E'\!-\!E_{n+l})\,\delta(E\!-\!E_{n+m})\,c_{n,n+l}\right>\\
&=&\displaystyle L^d\int \!\!d^dr\,  \Bigl<\rho(E'',\bbox{r})
\rho(E',\bbox{r})\rho(E)\Bigr>
\equiv L^d\int \!\!d^dr\,  {\cal Q}(E'',E',E;\,\bbox{r})\,\label{Q2}
 \end{eqnarray}
  \end{mathletters}
where%
 \begin{eqnarray}
 \label{LDOS}
 \rho(E',\bbox{r})\equiv \sum_{n}
 \bigl|\psi_n(\bbox{r})\bigr|^2 \delta(E\!-\!E_{n})
 \end{eqnarray}
is the local density of states (LDoS). Thus the correlation function 
involves, by definition, correlations of eigenvalues and eigenstates. 
By the assumption of homogeneity of energy space we see that
${\cal Q}(E'',E',E)$ is a function only of energy differences
 $ \omega=E\!-\!E''$
and $ \omega'=E'\!-\!E''$, so that we can put $E''=0$ 
without loss of generality.
Integrating Eq.\ (\ref{Q1}) over $E''$ prior and after putting $E''$
to $0$ then yields the following identity:
\begin{eqnarray*}
{\cal{N}}Q(\omega ',\omega )=E_0{\cal Q}(0, \omega ', \omega ) 
\end{eqnarray*}
where $E_0$ is energy bandwidth and ${\cal{N}}$ is total number of
energy levels.  From this identity and Eq.\ (\ref{QEE}) we obtain
the following representation for $Q(\omega ',\omega )$:
\begin{eqnarray}
\label{G}
Q(\omega ',\omega )={\Delta\over L^d}\left< \sum_{n,l,m}
\delta(E_n) \,\delta(E_{n+l}\!-\!E_n \!-\!  \omega ') \,
\delta(E_{n+m}\!-\!E_n\!-\! \omega )
\,c_{n,n+l} \right>\equiv \Delta{\cal Q}(0, \omega ', \omega )\,.
\end{eqnarray}

We now apply the Kirkwood approximation\cite{Kirkwood} to the correlation 
function, Eq.\ (\ref{Q2}), in the following form:
\begin{eqnarray}
\label{Kirk}
{\cal Q}(0,\omega ',\omega ;\,\bbox{r})=\frac{\Bigl<\rho(0,\bbox{r})
\rho(\omega ',\bbox{r})\Bigr>
\Bigl<\rho(0,\bbox{r})\rho(\omega )\Bigr>
\Bigl<\rho(\omega ',\bbox{r})\rho(\omega )\Bigr>}{\rho^3}
+B(0,\omega ',\omega ;\,\bbox{r})
\end{eqnarray}
where the numerator is chosen to incorporate all pairwise correlations,
and the denominator ensures the correct limiting value as
each energy difference tend to infinity. The correction $B$ is
small if the approximation is good, which is expected to be the case
unless $E''$, $E'\equiv E''\!+\!\omega'  $ and 
$E\equiv E''\!+\!\omega$ are all close together. Now from the
definitions of $\rho(E)$ and $\rho(E,r)$, Eqs.\ (\ref{rho})
and (\ref{LDOS}), the averages in
Eq.\ (\ref{Kirk}) can be expressed via TLCF,  $R(\omega)$,
\begin{eqnarray*}
\Bigl<\rho(E+\omega,\bbox{r})\,\rho(E)\Bigr>=
\Bigl<\rho(E+\omega)\,\rho(E)\Bigr>
=\rho^2\left(1+R(\omega)\right)\,,
\end{eqnarray*}
and the Fourier transform of the return probability, $p(\omega)$,
\begin{eqnarray*}
\int \!\!d^dr\, \Bigl<\rho(E+\omega,\bbox{r})\,\rho(E,\bbox{r})\Bigr>=
\frac{1}{2\pi\hbar\Delta}\,\left[p(\omega)+2\pi\hbar \rho\right]
={\rho\over\Delta}c(\omega)\,,
\end{eqnarray*}
where we have taken into account that $p(\omega)$ is expressed via irreducible
part of the LDoS correlation function, and used the relation (\ref{pC}).

Thus the Kirkwood approximation of Eq.\ (\ref{Kirk}) yields the expression
\begin{eqnarray}
\label{Kirk1}
\begin{array}{rcl}
Q(\omega ',\omega )&=&\displaystyle {\rho\over\Delta}
\Bigl\{
c(\omega ')\left[1+R(\omega )+R(\omega '\!-\!\omega )\right]\Bigr\}\\
&+&\displaystyle{\rho\over\Delta}
c(\omega ')R(\omega )
R(\omega '\!-\!\omega )+
\Delta\int \!\!d^dr\, B(0,\omega ',\omega ;\bbox{r})\,. 
\end{array}
\end{eqnarray}
The first line of the above equation represents the terms included
in the Brownian motion formula, and the second line is the
correction. We can now rewrite the exact relation Eq.\ (\ref{p=Q})
using Eq.\ (\ref{Q}) and the factorization given by the 
first line of  Eq.\ (\ref{Kirk1}) to
obtain
\begin{eqnarray}
\label{Kfac1}
K(t)=\left(\frac{1}{4\pi\hbar\rho}\right)
\frac{tp(t)-A(t)}{\int_0^t dt'\,C(t')}
\end{eqnarray}
where the remainder $A(t)$ is given by
the Fourier transform in the form (\ref{QFT}) of the second line 
of Eq.\ (\ref{Kirk1}).  Note that 
in deriving Eq.\ (\ref{Kfac1}), we, as above, have neglected a non-integrated
$\delta(t)$ function which does not contribute to the final result. 

 If we ignore the remainder $A(t)$,
 Eq.\  (\ref{Kfac1}) is equivalent to Eq.\ (\ref{Kfac}) and thus to
 the Brownian motion result, Eq.\ (\ref{K1})
We see that the use of the
Kirkwood approximation is just a more formal way of doing the
factorization discussed previously. The factor of $2$ that occurs
in Eq.\ (\ref{Qfac}) emerges naturally, as the two equivalent two-level
correlations are automatically taken into account. 
It is obvious that the second line of  Eq.\ (\ref{Kirk1}) 
could be small only if  $|\omega|,\,|\omega'|,\,|\omega-\omega'|\gg
\Delta$. Thus we could  expect $A(t)$ to be small, and 
 Eq.\ (\ref{K1}) to be valid only outside of
 the quantum regime, i.e.\ for  $t\alt t_{\!_H}$. 

What we would like to be able to do is to find out how big the remainder
term $A(t)$ actually is in this non-quantum limit.
We can see that $A(t)$ consists of two
parts:-- the first involves a product of three two-level correlation
functions, and is the Kirkwood approximation's attempt to represent
three-level correlations in terms of two-level correlations; the second
is the correction to the approximation itself. To proceed we will
introduce diagrammatic perturbation theory in the next section,
rewrite our factorization in this language, and hence 
discover what is left over.

\section{Diagrammatic Analysis}
Our aim now is to check the validity of Eq.\ (\ref{K1}) beyond the
trivial diagonal approximation in which one can just neglect 
the $t$-dependence of the denominator,
and substitute the classical limit of $p(t)$ into the numerator of this
expression. The simplest way of doing this is to calculate
both $K(t)$ and $p(t)$ to higher order in perturbation theory,
and to compare directly the r.h.s.\ and l.h.s.\ of  Eq.\ (\ref{K1}).
It is more instructive, however, to analyze diagrammatically
the exact relation (\ref{p=Q}) and the decoupling procedure based on
Eqs.\ (\ref{Q}) and (\ref{Kirk1}).
In this way we will not only check the accuracy of Eq.\ (\ref{K1}),
but arrive at an alternative factorization given by Eq.\ (\ref{K4}).
 
Let us note that the role of higher order corrections is different in
$d=2$ and $d>2$. In the $2d$ case they are universal  
in the sense that
they are due to diffusive motion of electrons throughout the whole
sample and almost insensitive to details of motion at the
ballistic scale. In $d>2$, corrections are mainly due to the motion
at the ballistic scale, and proportional to powers of an additional
small parameter $(t/t_{\text{el}})^{d/2-1}$ (as well as to powers of the
standard weak disorder parameter -- inverse dimensionless conductance).
These corrections are not only small but of no particular interest as
they do not drive the system from weak to strong-disorder regime.
In contrast to this, the $2d$ corrections do describe crossover from the weak
to strong disorder, and are widely believed to be more relevant for the
vicinity of the metal-insulator transition for $d>2$ than those calculated
in the metallic limit directly in $d>2$ dimensions. Moreover, as
in the diagonal approximation
$K(t)\propto t p(t) \sim \text{const}$ for $t\alt t_{\text{erg}}$ at $d=2$,
the TLCF vanishes in this approximation in the
diffusive regime ($E\agt E_c$). Therefore, in this regime the first
non-vanishing higher-order contribution governs the main effect rather than 
describing some correction. For all these reasons, we will consider mainly 
the  $2d$ case in this section. 

\subsection{General Relations}
To be able to work in complete generality we will rewrite the 
exact equation (\ref{p=Q})
in terms of Green's  functions that can then be expanded
using standard diagrammatic methods. The form of the diagrams for
$K(t)$ is by now well known, but those for $p(t)$ and the three-level
correlator $Q(t)$ are less familiar.

We start with the standard expression for DoS in terms
of exact Green's  functions where, as well as elsewhere in this section,
we use the units $\hbar=1$:
\begin{eqnarray}
\label{GrGa}
\rho(E,\bbox{r})=\frac{i}{2\pi}\left[G^R(E;\bbox{r},\bbox{r})-G^{A}
(E;\bbox{r},\bbox{r})\right]\,,
\end{eqnarray}
where the retarded, $G^R$ and advanced $G^A$ Green's  functions are
defined by
\begin{eqnarray}
\label{Gdef}
G^{R,A}(E;\bbox{r},\bbox{r}')=
\sum_n\frac{\psi_n(\bbox{r})\psi_n(\bbox{r}')}{E-E_n\pm i0}\,.
\end{eqnarray}
For the density-density correlation function
 we get the formula
\begin{eqnarray}
\label{rhorho}
R(\omega)=
\left(\frac{i\Delta}{2\pi }\right)^2\int \!\!d^dr  \int \!\!d^dr\, '
\Bigl<{\cal G}(0;\bbox{r},\bbox{r})
{\cal G}(\omega;\bbox{r}',\bbox{r}')\Bigr>\,,
\end{eqnarray}
where ${\cal G}=G^R-G^A$. The diagrams for this then consist of two
loops, each with an external vertex (with coordinates
 $\bbox{r}$ and $\bbox{r}'$), as
shown in Fig.\ (4a). Averaging over the disorder ensemble then leads
to the presence of impurity lines both within a loops and across loops.
\begin{figure}
\epsfxsize=0.8\textwidth
\epsfysize=0.8\textheight
\hspace*{\fill}\epsffile{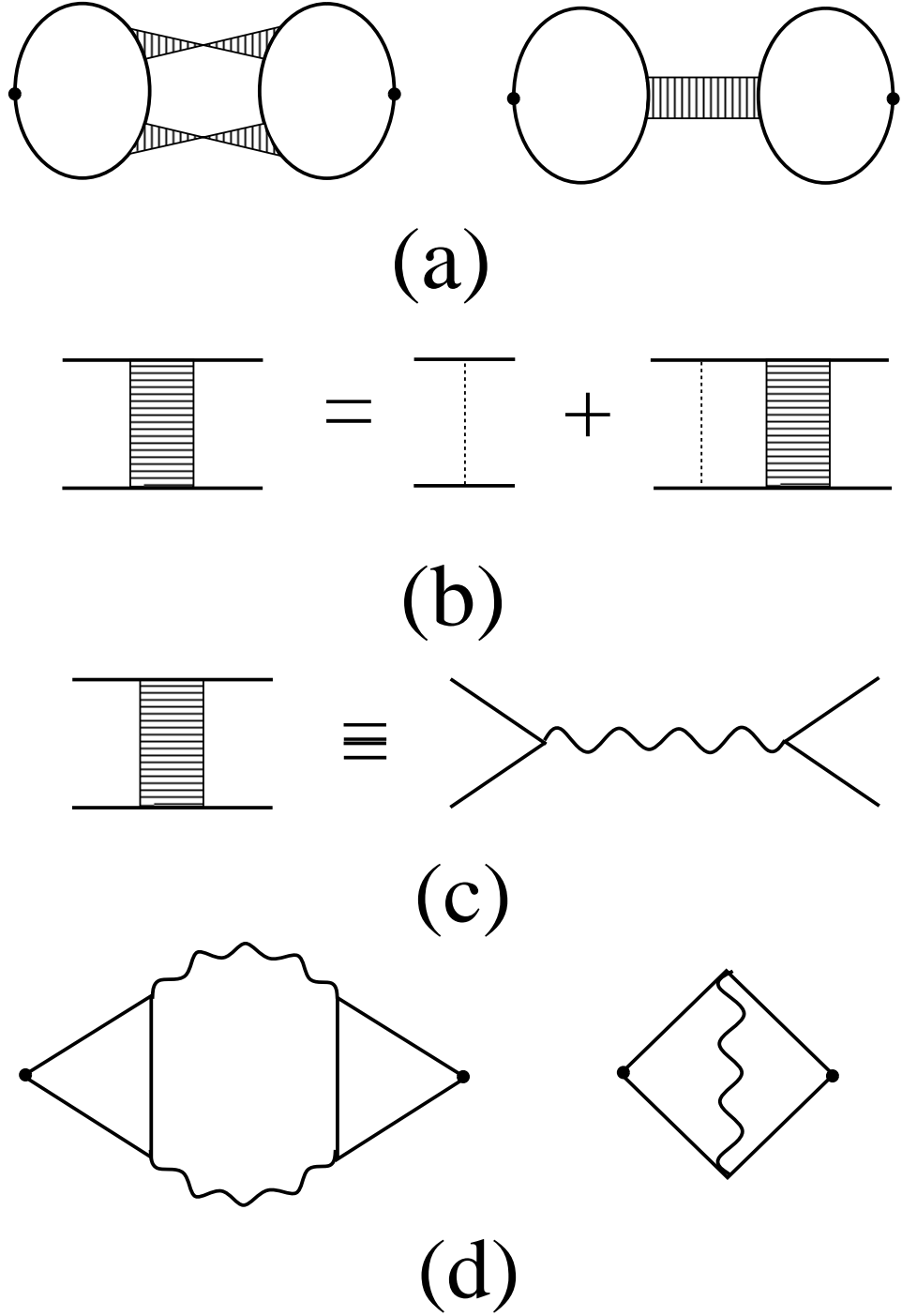}\hspace*{\fill}
\vspace{0.3truein}
\caption{(a) The two one-loop diagrams for $R(\omega)$ in a disordered
metal. The diagram on the left gives the (dominant) Altshuler-Shklovskii
contribution; that on the right is smaller by a factor $\Delta\tau$,
and is usually ignored. (b) Definition of the impurity ladders occuring
in (a). (c) Rewriting the impurity ladder as an effective propagator.
(d) Rewriting the diagrams in (a) using the notation of (c). The
electron Green's  function lines end up in the so-called Hikami boxes.}
\end{figure}
Only  the connected diagrams
contribute to $R(\omega)$, while the trivial unconnected diagrams
are cancelled by $-1$ in the definition of $R(\omega)$,  Eq.\ (\ref{R(E)}).

Similarly, $p(\omega)$ is given by
\begin{eqnarray}
\label{pdiag} 
p(\omega)&=&\frac{2\pi}{\rho L^d}\int \!\!d^dr\, 
\left<\rho(E+\omega,\bbox{r})\rho(E,\bbox{r})\right>
=-\frac{\Delta}{2\pi}\int \!\!d^dr\, 
\left<{\cal G}(\omega;\bbox{r},\bbox{r}){\cal G}(0;\bbox{r},\bbox{r})\right>
\,. 
\end{eqnarray}
\begin{figure}
\epsfxsize=6truein
\hspace*{\fill}\epsffile{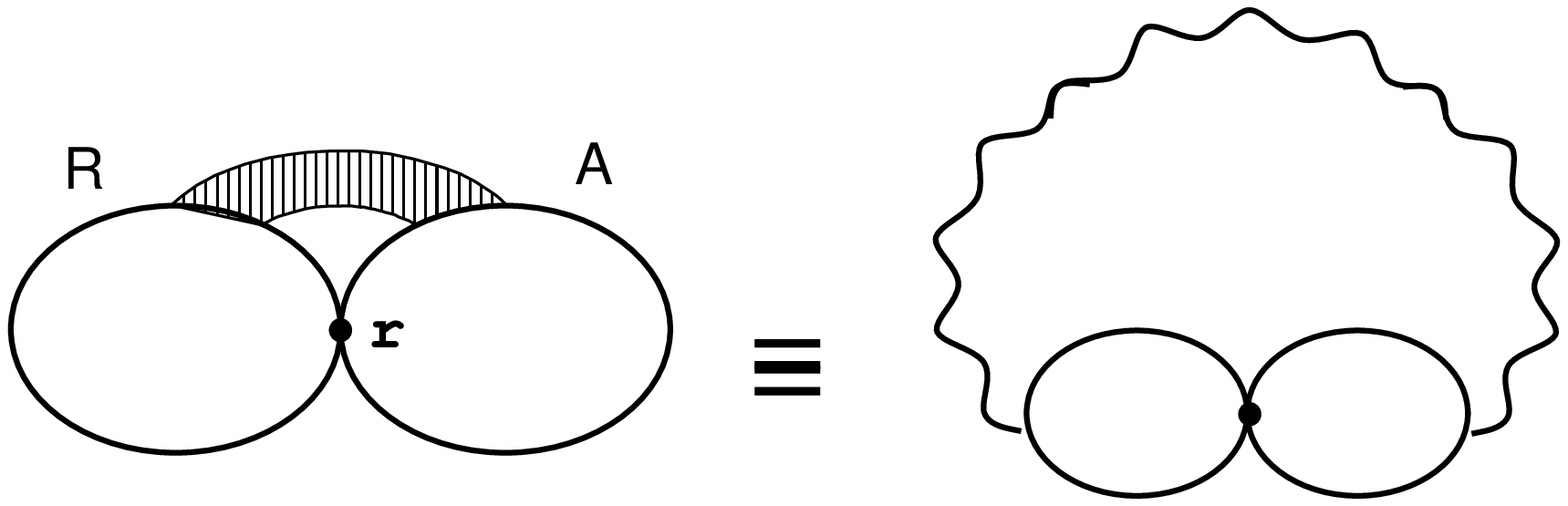}\hspace*{\fill}
\vspace{0.5truein}
\caption{The one-loop
diagram for the quantum return probability $p(\omega)$ in the ladder
and Hikami-box representations.}
\end{figure}
The difference between the formulae (\ref{rhorho}) and (\ref{pdiag})
is that in the latter the densities of states are evaluated at the same point
$\bbox{r}$ in space rather than at different points. In diagrammatic terms
this means that $p(\omega)$ consists of two loops  connected to a single
external point $\bbox{r}$, as shown in Fig.\ (5).

In the diffusive regime of the
disordered metal it is impurity ladders describing diffusive motion
which give the important energy
dependent contributions. They must have both a $\scriptsize 
R$ and $\scriptsize A$ line --
so only $\scriptsize RA$ and 
$\scriptsize AR$ diagrams can contribute. Since the latter are
complex conjugates of each other it follows that we can write
the expressions for $R(\omega)$ and $p(\omega)$ in the following form:
\begin{eqnarray}
\label{Kdiag}
R(\omega)&=&
\frac{\Delta^2}{2\pi^2}\,\Re \text{e}
\int \!\!d^dr \int \!\!d^dr\, '
\left<G^{R}(\omega;\bbox{r},\bbox{r})G^{A}(0;\bbox{r}',\bbox{r}')\right>\,,
\\
\label{pdiag1}
p(\omega)&=&\frac{\Delta}{\pi}\,
\Re \text{e}
\int \!\!d^dr\, 
 \left<G^{R}(\omega;\bbox{r},\bbox{r})G^{A}(0;\bbox{r},\bbox{r})\right>\,.
\end{eqnarray}
The impurity ladders (Fig.\ 4b) which connect the two loops
describe either diffuson or Cooperon propagators
given in momentum space by
\begin{eqnarray}
\label{ladder}
D(q;\omega)=\frac{1}{2\pi\rho t_{el}^2}\,
\frac{1}{Dq^2-i\omega}\,.
\end{eqnarray}
We will consider both the cases with and without time-reversal symmetry, 
referring to them as the orthogonal case ($\beta=1$) and
unitary case ($\beta=2$). Diagrammatically, the latter differs from the
former by the absence of the Cooperon contributions which are time-reversed
impurity ladders. 

The loops could be connected by arbitrary number of the ladders. 
In the lowest order there are the  two contributions to $R(\omega)$
shown in Fig.\ 4a, and one contribution to $p(\omega) $
(Fig.\ 5).
The dominant contribution to  $R(\omega)$ -- which is called
Altshuler-Shklovskii diagram --
has two impurity ladders; the diagram with only one ladder gives
much smaller contribution.

Thus the perturbation order of a diagram is not determined by the number of
ladders. A standard way  \cite{GLKh,Hik:81}
to determine this order, and to make
the calculation of diagrams more straightforward is rewriting them in the
form  where impurity ladders are represented as 
propagators (a wavy line in Fig.\ 4c); this is convenient 
since the ladders involve small momentum, $q\ell\ll 1$, and energy, 
$\omega t_{el}\ll 1$.
All other Green's  function lines are  absorbed into effective
interactions between propagators (as shown in Figs.\ 4 and 5)
known as Hikami boxes.
Then the order of a
diagram  is  the number of independent momenta occuring in
the propagators, which is just the 
number of loops made of the wavy lines. (Each box corresponds to a
single spatial point $\bbox{r}$ and can be thought of as being 
contracted into a point; thus  any diagram  
would consist of some number of wavy loops).
The loop-expansion parameter here is $1/g$ where $g$ is the
dimensionless conductance of the sample.

The one-loop diagrams for $R(\omega)$
 are shown in Fig.\ 4d. 
The two external vertices may be in
separate (odd) boxes, or the same (even) box; we ignore the
latter since they are smaller than the former by factor 
$\Delta t_{el}\sim (\ell/L)^d$ in the
same order of perturbation theory. The remaining Hikami boxes that
do not contain an external vertex have an even number of sides.

Rewriting diagrams for $p(\omega)$
in terms of Hikami boxes  we find that the external vertex (corresponding
to the point $\bbox{r}$ in Eq.\ (\ref{pdiag1}))
has two boxes connected to it -- these can either
both be odd boxes, or both be ``2-gons'' (larger even boxes are
not allowed since one can always string impurity ladders across them to
recover the ``2-gon'' structure). All other boxes have no external
vertex and so have an even number of sides. Note that all the Hikami 
boxes must be ``dressed'', i.e.\ single-impurity lines should 
connect (without mutual intersection)
those {\it non-adjacent} sides of the boxes which have the same analyticity 
(both $\scriptsize R$, or both  $\scriptsize A$)\cite{Hik:81}.
In all diagrams below the boxes are assumed to be dressed.
The one-loop contribution to
$p(\omega)$ is given in Fig.\ 5 in both the ladder and Hikami representations.
All the two-loop contributions to $p(\omega)$ and $R(\omega)$ are given 
in the Hikami representation in Fig.\ 6. 
\begin{figure}
\epsfxsize=0.8\textwidth
\hspace*{\fill} \epsffile{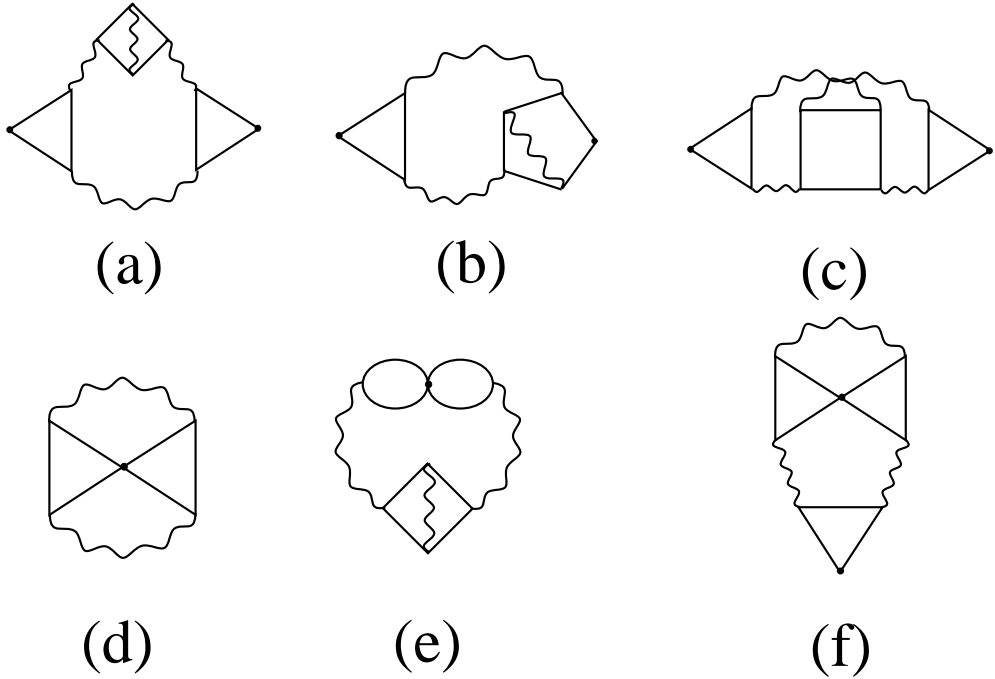}\hspace*{\fill}
\vspace{0.5truein}
\caption{Two loop order diagrams to: $R(\omega)$ (a, b, c);
$\widetilde p(\omega)$ (d, e); and $\widetilde
Q(\omega)$ (f). All diagrams contribute in
the orthogonal case, while only (d) and (f) contribute in the unitary
case.}
\end{figure}

Before comparing the higher-order contributions made to Eq.\ (\ref{K1}) by 
the diagrams for $R(\omega)$ and $p(\omega)$, we
consider the three-level correlator $Q(\omega',\omega)$. Starting from
Eq.\ (\ref{G}), and rewriting the densities of states via 
Eq.\ (\ref{GrGa}) we get
\begin{eqnarray}
\label{Qdiag}
Q(\omega',\omega)=\Delta \left(\frac{i}{2\pi}\right)^3 
\int \!\!d^dr\,  \int \!\!d^dr\, '
\Bigl<{\cal G}(0;\bbox{r},\bbox{r})
{\cal G}(\omega';\bbox{r},\bbox{r}){\cal G}(\omega;\bbox{r}',\bbox{r}')\Bigr>
\end{eqnarray}
Following the procedure used for $R(\omega)$ and $p(\omega)$ above we see that
in the ladder representation
the diagrams for $Q(\omega',\omega)$ consist of three loops, 
two of which are joined at
the external vertex $\bbox{r}$, and the third having the external vertex
$\bbox{r}'$.
There are now several classes of diagrams that are not fully connected,
as shown in Fig. (7a). (The shaded strips there include symbolically all
possible combinations of impurity ladders). The most trivial has all
loops unconnected and yields a constant term. The two others are 
reducible as they
have only two out of the three loops connected by impurity
ladders, and contribute terms proportional to
$\rho^3 R(\omega)$, $\rho^3R(\omega'-\omega)$ and $\rho p(\omega')$.
These reducible contributions  add up to give
\begin{eqnarray}
\label{Quc}
Q_{uc}(\omega',\omega)=\left(\frac{\rho L^d}{2\pi}\right)\left\{
2\pi\rho[1+R(\omega)+R(\omega'\!-\!\omega)]+p(\omega')\right\}
\end{eqnarray}
\begin{figure}
\epsfxsize=0.8\textwidth
\hspace*{\fill}\epsffile{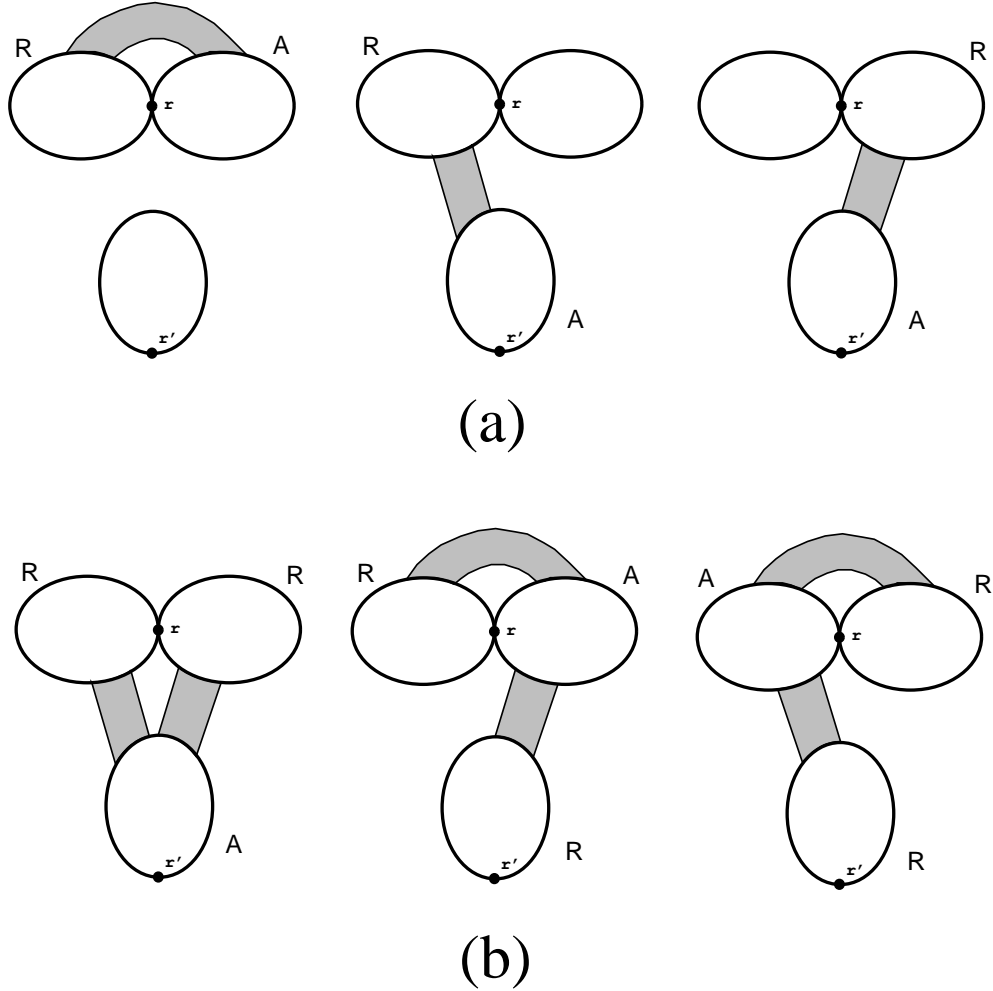}\hspace*{\fill}
\vspace{0.5truein}
\caption{(a) The reducible contributions to the three-level
correlator $Q(\omega',\omega)$. The first diagram yields no contribution
to the function $Q(t)$, whilst the other two yield equal contributions
to $Q(t)$ that are proportional to $K(t)$. (b) The irreducible
contributions to the three-level correlator $Q(\omega',\omega)$.
For analyticity reasons the last two diagrams give no contribution to the
function $Q(t)$, so we need consider only diagrams of the first type.}
\end{figure}
We see that the r.h.s.\  of the above is similar , but not identical to,
the first line of the r.h.s.\  of Eq.\ (\ref{Kirk1}). The difference is that
the is no
$p(\omega')$ multiplying the $R(\omega)$ and $R(\omega'\!-\!\omega)$ in 
Eq.\ (\ref{Quc}), only its constant part $2\pi\rho$. If we substitute 
this term into Eq.\ (\ref{p=Q}) we obtain
$K(t)=(2\pi\hbar)^{-1}t p(t)$.
In other words, the reducible diagrams of the three-level correlator
$Q$ are enough to reproduce the diagonal approximation of semi-classics.
In order to recover our formula, Eq.\ (\ref{K1}), we will need to
analyze the irreducible contributions to $Q$ given schematically in Fig.\
7b.

If we look at the irreducible contributions the first thing to note
is that each diagram can only have ladders between two of the three
{\it pairs} of loops. 
This is because ladders are always between an
$\scriptsize R$ and an $\scriptsize A$
line, and so in an irreducible diagram either two loops are 
$\scriptsize R$ and the
third $\scriptsize A$ or vice versa. There are therefore 
6 possibilities which
come in complex conjugate pairs: $\scriptsize AAR$ and 
$\scriptsize RRA$; $\scriptsize ARA$ and $\scriptsize RAR$;
$\scriptsize RAA$ and $\scriptsize ARR$. The irreducible part of 
$Q(\omega',\omega)$ can then be
written in the form
\begin{eqnarray}
\label{Qc}
\begin{array}{c}
\displaystyle
Q_{c}(\omega',\omega)=\frac{1}{4\pi^3\rho L^d}
\int \!\!d^dr\int \!\!d^dr\, ' \Im {\text {m}}\left\{
\left<G^A(0;\bbox{r},\bbox{r})G^A(\omega;\bbox{r},\bbox{r})
G^R(\omega';\bbox{r}',\bbox{r}')\right>\right.\\
\displaystyle
\left.+\left<G^A(0;\bbox{r},\bbox{r})G^R(\omega;\bbox{r},\bbox{r})
G^A(\omega';\bbox{r}',\bbox{r}')\right>
+\left<G^R(0;\bbox{r},\bbox{r})G^A(\omega;\bbox{r},\bbox{r})
G^A(\omega';\bbox{r}',\bbox{r}')\right>\right\}
\end{array}
\end{eqnarray}
When rewritten in the Hikami-box representation the
diagrams for $Q$ consist of the two boxes connected to external vertex
$\bbox{r}$ found in $p(\omega)$ diagrams, plus a single odd box connected to
external point $\bbox{r}'$ as found in $R(\omega)$ diagrams; these are then 
held
together by wavy lines  and even boxes with no external vertices. $Q_{AAR}$ 
diagrams cannot have the ``2-gon'' structure in the part connected to
external point $\bbox{r}$ since such a structure ends in ladders, which cannot
happen here as both lines of the ladder would be 
$\scriptsize A$. $Q_{ARA}$ and
$Q_{RAA}$ diagrams can have this ``2-gon'' structure.

If we are to re-derive our main result Eq.\ (\ref{K1}) diagrammatically
we will need to show that the connected part of $Q(\omega',\omega)$ factors
into a product of $R$ and $p$. This can be seen by recalling the
derivation of Eq.\ (\ref{K1}) via the Kirkwood approximation, where
such factorization occurs in the first line of Eq.\ (\ref{Kirk1}).
The fact that only two pairs of loops can have
ladders between them suggests a possible reason for this to occur.
At this point we note that we are interested not in $Q(\omega',\omega)$ 
itself
but rather in $Q(t)$ which is derived from it by Fourier transforms as
in Eq.\ (\ref{QFT}). The analyticity properties of $Q(\omega',\omega)$ 
diagrams
will lead to some types giving no contribution to $Q(t)$.
Eventually, a factorization emerges from
this analysis which is different to that
suggested in Eq.\ (\ref{Kirk1}), but
which, remarkably,
yields the same spectral form factor in two dimensions, to third-loop order.

\subsection{Analytical properties of the three-level correlation function}

First let us prove that only the $Q_{AAR}$ diagrams survive, starting 
with the formula for $Q(t)$ given by Eq.\ (\ref{Q}),
\begin{eqnarray}
\label{tQ(t)}
\begin{array}{rcl}
Q(t)&=&\displaystyle
\int_0^t dt'\int
d\omega'\int d\omega e^{-i\omega't'}e^{i\omega t}Q(\omega',\omega)\\
&=&\displaystyle\int
d\omega'\int d\omega
{e^{i\omega t}-e^{i(\omega-\omega')t}\over i\omega'}Q(\omega',\omega)
\end{array}
\end{eqnarray}
Taking the Fourier transform of $Q(t)$ then yields
\begin{eqnarray}
\label{tQ(t)FT}
\begin{array}{rl}
&\displaystyle \int dt e^{i\overline{\omega}t}\int d\omega'\int d\omega
\frac{e^{i\omega t}-e^{i(\omega-\omega')t}}{i\omega'}Q(\omega',\omega)\\
= 2\pi &\displaystyle \int d\omega'\int d\omega \frac{1}{i\omega'}\left[
\delta(\overline{\omega}+\omega)-\delta(\overline{\omega}+\omega-\omega')
\right]Q(\omega',\omega)\\
= -2\pi i&\displaystyle \int \frac{d\omega'}{\omega'}\left[
Q(\omega',-\overline{\omega})-Q(\omega',\omega'-\overline{\omega})\right]
\end{array}
\end{eqnarray}
Now consider the analyticity properties of the $\scriptsize ARR$,
 $\scriptsize ARA$ and $\scriptsize AAR$
respectively. They can be written in the form
\begin{eqnarray}
Q_{RAA}(\omega',\omega)&=&f(-\omega',-\omega)\nonumber\\ 
\label{Anal}
Q_{ARA}(\omega',\omega)&=&f(\omega',\omega'-\omega)\\
Q_{AAR}(\omega',\omega)&=&f(\omega,\omega-\omega') \nonumber\,.
\end{eqnarray}
where in each case $f(\omega_1,\omega_2)$ is a function analytic in the 
u.h.p.\ 
for both arguments. For $Q_{RAA}$ Eq.\ (\ref{tQ(t)FT}) gives
\begin{eqnarray}
\label{Q+--}
-2\pi i\int {d\omega'\over \omega'}\left[f(-\omega',\overline{\omega})
-f(-\omega',-\omega'+\overline{\omega})\right]
=-2\pi^2\left[f(0,-\overline{\omega})-f(0,-\overline{\omega})\right]=0\,,
\end{eqnarray}
since upon closing each term in l.h.p. we only get contributions from the
pole at $\omega'=0$, and these cancel.
Similarly for $Q_{ARA}$ we get upon closing in the u.h.p.
\begin{eqnarray}
\label{Q-+-}
-2\pi i\int {d\omega'\over \omega'}\left[f(\omega',\omega'+\overline{\omega}) 
-f(\omega',\overline{\omega})\right]
=2\pi^2\left[f(0,\overline{\omega})-f(0,\overline{\omega})\right]=0\,.
\end{eqnarray}
Finally for $Q_{AAR}$ we get
\begin{eqnarray}
\label{Q--+}
-2\pi i\int {d\omega'\over \omega'}\left[f(-\overline{\omega},-\omega'-
\overline{\omega})  
-f(\omega'-\overline{\omega},-\overline{\omega})\right]=-(2\pi)^2 
f(-\overline{\omega},-\overline{\omega})\,.
\end{eqnarray}
where we closed first term in l.h.p., second in u.h.p. to yield 
contributions that add up. The above term can be rewritten in the
form $-(2\pi)^2 Q_{AAR}(\omega'=0,-\overline{\omega})$, so it follows that
diagrams for $Q_{AAR}$ can yield Fourier transform of $Q(t)$ directly. 
We have therefore verified our previous assertion, and moreover have
derived the contribution to the Fourier transform of $Q(t)$ coming
from the $Q_{AAR}$ diagrams. Diagrammatically the above means that all
ladders in $Q_{AAR}$ diagrams will have the same energy dependence.

Let us discuss what the above means for our factorization hypothesis.
The reason we expected that it would be the $Q_{RAR}$ and $Q_{ARR}$
diagrams that survived is that we can envisage a natural factorization
into terms with energy dependence of the form $R(\omega)p(\omega')$ and 
$R(\omega'-\omega)p(\omega')$, which are exactly what is needed to reproduce 
Eq.\ (\ref{K1}). The fact that it is $Q_{AAR}$ which survives means
that our factorization hypothesis must be altered to have an energy
dependence $R(\omega)p(\omega)$, which leads to a convolution in $t$-space.
More precisely, we expect the $ \omega$-space factorization to have the
form
\begin{eqnarray}%
\label{twofac}
\widetilde Q(\omega)
=-\frac{(2\pi)^2i}{\Delta}\,\widetilde {p}(\omega)\,\widetilde {R}(\omega)\,.
\end{eqnarray}
Here we have introduced the analytical functions $
\widetilde Q(\omega),\,\widetilde {p}(\omega)$ and
$\widetilde {R}(\omega)$ as follows:
\begin{eqnarray}%
\label{tilde}
\begin{array}{rcl}
\widetilde Q(\omega)&\equiv & Q_{AAR}(0,\,\omega)\,;\\
{p}(\omega)&\equiv & 2\Re\text{e}\,\widetilde{p}(\omega)\,;\\
{R}(\omega)&\equiv & 2\Re\text{e}\,\widetilde{R}(\omega)\,.
\end{array}
\end{eqnarray}
Note that diagrammatically we calculate exactly these analytical functions.
Before giving the results for higher-loop contributions, note that 
calculations of $\widetilde{R}(\omega)$ are considerably simplified due to
the fact that it can be obtained
as the second derivative of ``free energy'' $F(\omega)$:
\begin{eqnarray}
\label{Fdef}
\widetilde{R}(\omega)=-\left({\Delta\over 2\pi}\right)^2
\frac{\partial^2}{\partial \omega^2}\widetilde F(\omega)
\end{eqnarray}
Here $F(\omega)$ is given by the sum of all 
diagrams that have no external vertices, and
 the coefficient of proportionality
is chosen so as to make $F(\omega)$ dimensionless.

\subsection{Calculation of diagrams up to three-loop order}

As a result of the above discussion we can now lay out the diagrammatic
program ahead. First, we will calculate the contributions to 
$\widetilde F(\omega)$,
$\widetilde p(\omega)$, and $\widetilde Q(\omega)$ up to three-loop order. 
Then we will first verify 
Eq.\ (\ref{p=Q}) (which has been obtained assuming spectral
homogeneity), and then check the factorization scheme
of Eq.\ (\ref{twofac}) which is dictated by the
analytical structure described above. This factorization is equivalent to
that given by Eq.\ (\ref{K4}).  Finally, we will show how, having verified
this factorization up to the three-loop order, to check the factorization
that comes out of the Kirkwood approximation, Eq.\ (\ref{Kirk1}),
 and is equivalent to the result of the Brownian-motion model, Eq.\ (\ref{K1}). 
We will show that up to the 
third-loop order both the factorizations are identical and exact.  Therefore, 
Eq.\ (\ref{K1}) obtained within the Brownian motion picture turn out 
to be correct to quite a nontrivial order of perturbation theory. 

Let us introduce a notation that is convenient for further analysis:
\begin{eqnarray}
\label{PP}
\PP_{1\ldots n}\equiv D(\bbox{q}_1+\ldots +\bbox{q}_n)^2 -i\omega
\end{eqnarray}
In particular, $\PP_1\equiv Dq_1^2-i\omega$.

Now,  two-loop order contributions corresponding to the diagrams
in Fig.\ 8(a) (equivalent to those in Fig.\ 6a--c) and
in Fig.\ 6(d--f) may be written as
\begin{eqnarray}
\label{twoloop}\nonumber
\widetilde F_2(\omega)&=&\left({2-\beta\over \beta}\right)
{\Delta\over2\pi}
\sum_{q_1,q_2}\frac{\PP_1+\PP_2 +i\omega}{\PP_1\,\PP_2}\,;\\
\widetilde p_2(\omega)&=&\left({2\over\beta}\right)
{\Delta\over2\pi L^d} 
\Re\text{e}
\sum_{q_1,q_2}\left\{\frac{1}{\PP_1\,\PP_2}+
2(2-\beta)\frac{\PP_1+\PP_2 +i\omega}{\bigl(\PP_1
\bigr)^2\,\PP_2}
\right\}\,;\\
\widetilde Q_2(\omega)&=&-i\left(
{2\over \beta}\right)^2 {\Delta\over L^d}\sum_{q_1,q_2} \frac{1}
{\bigl(\PP_1 \bigr)^2\,\PP_2}\nonumber
\end{eqnarray}
Here we note the following. The diagrammatic approach could be used
both in the diffusive regime, $\omega\gg E_c$, where all the sums above
should be replaced by integrals, and in the ergodic regime, 
$E_c\gg\omega\gg\Delta$, where only contribution of zero mode (all 
$\bbox{q}=0$) survives. In the diffusive regime, regularization of
divergent integrals is required. Although we do not explicitly
calculate contributions of all diagrams below, in all algebraic manipulations
we use dimensional regularization near $d=2$. These manipulations
involve dealing with large $q\sim \ell^{-1}$, and so all diagrammatic
expressions we give here are not directly valid in the ergodic regime. 
However, it is straightforward to verify the accuracy of our factorization
scheme in the zero-mode regime as well.

The three-loop contributions 
to $F(\omega)$, $p(\omega)$,  and  $Q(\omega)$ are shown in Figs.\ 
8, 9, and 10, respectively.  In Fig.\ 8a we have also drawn the two-loop
contribution to  $F(\omega)$ which is equivalent to the three 
two-loop contributions to $R(\omega)$ shown in Fig.\ 6. Note that in 
the three-loop order there are 41 diagrams that contribute directly 
to $R(\omega)$, so that to have instead only the 5 diagrams for  $F(\omega)$,
as in  Fig.\ 8, is  a considerable simplification.
Nevertheless, these
three-loop results are quite bulky, and we list contributions of
\begin{figure}
\epsfxsize=0.8\textwidth
\hspace*{\fill} \epsffile{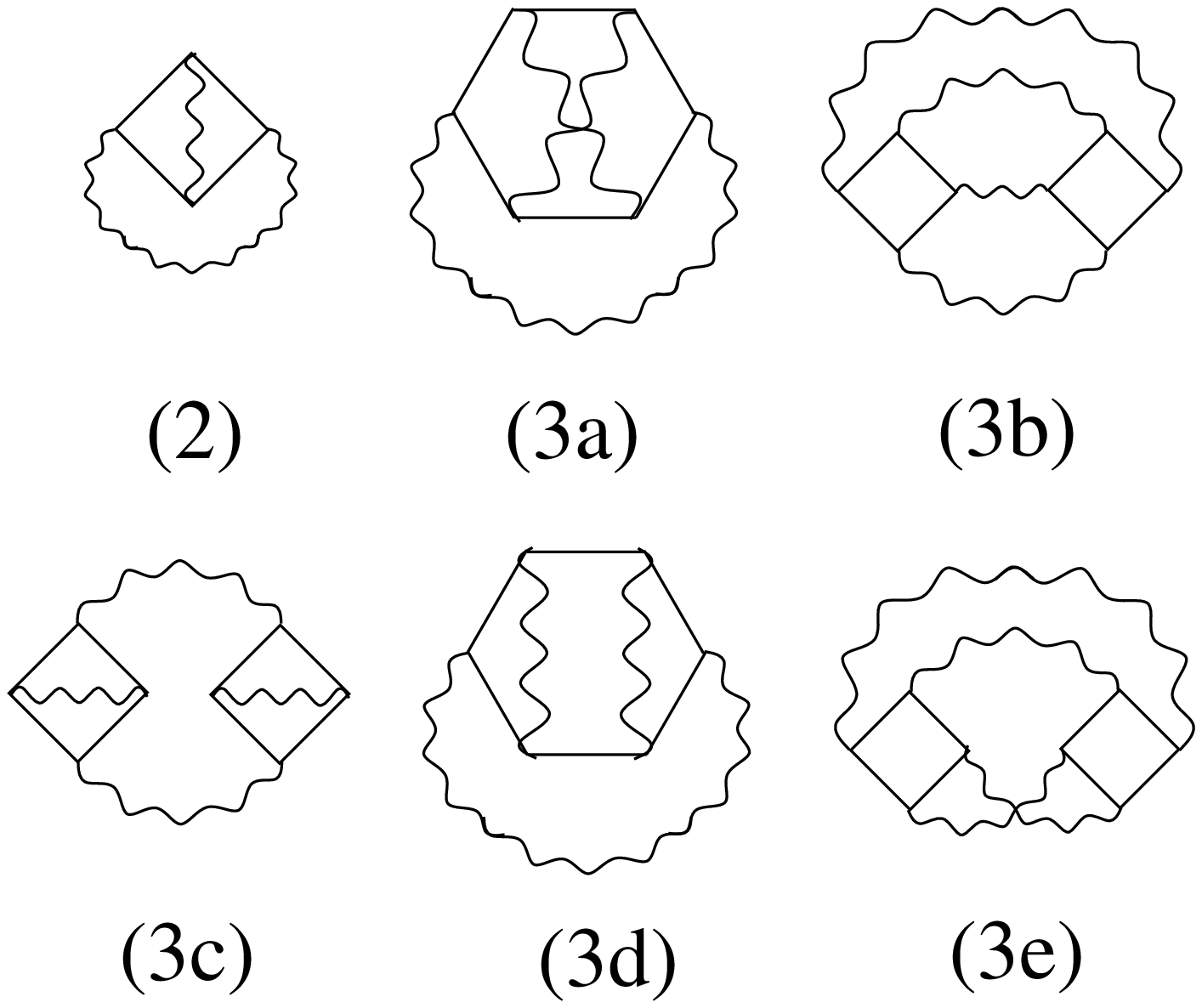}\hspace*{\fill}
\vspace{-1.9truein}
\caption{Two and three-loop order diagrams for the function $\widetilde
F(\omega)$.
The two-loop diagram is equivalent to those in Fig.\ 6(a--c), and contributes
only to the orthogonal case.
All 5 three-loop
diagrams contribute to the orthogonal case, while only
3a and 3b contribute to the unitary case.
}
\vspace{1cm}

\noindent
\epsfxsize=0.8\textwidth
\hspace*{\fill}\epsffile{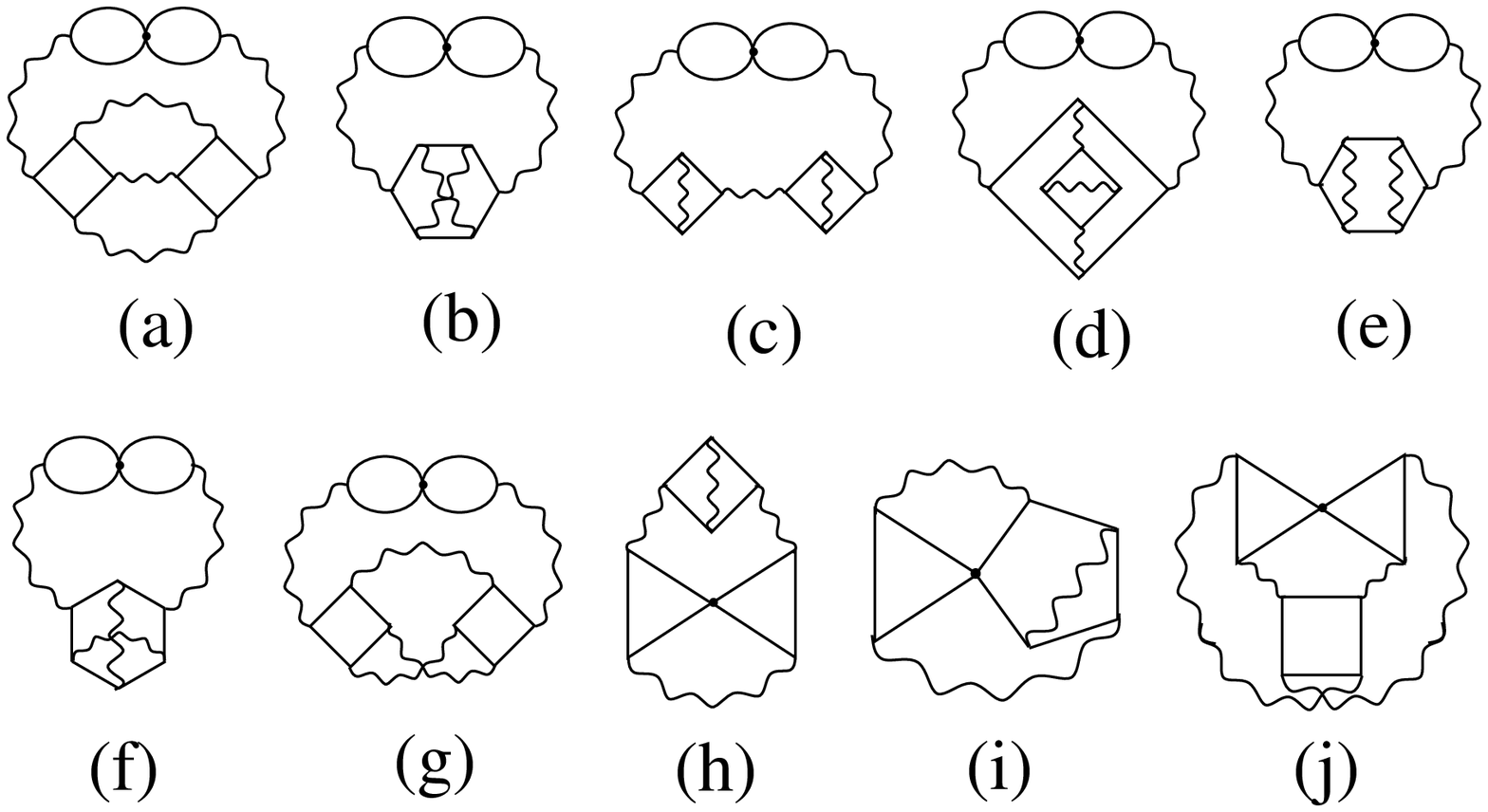}\hspace*{\fill}
\vspace{-2truein}
\caption{Three-loop order diagrams for the quantum return
probability $\widetilde p(\omega)$. All 10 diagrams contribute in the orthogonal
case, while only the first 2 contribute in the unitary case.}
\end{figure}
\begin{figure}
\epsfxsize=6truein
\epsffile{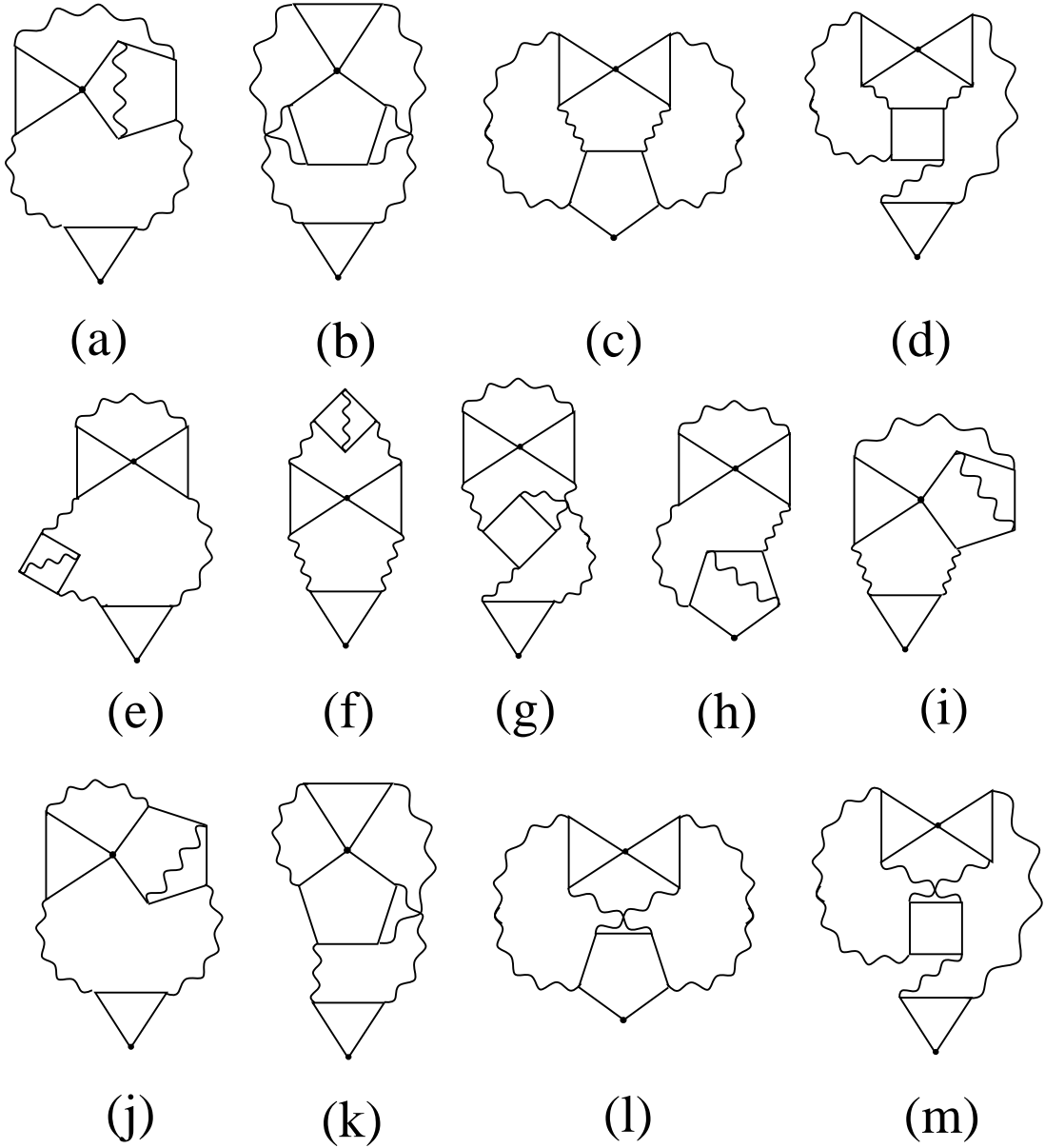}
\vspace{0.5truein}
\caption{Three loop order diagrams for the function $\widetilde
Q(\omega)$. All 13
diagrams contribute in the orthogonal case, whilst only the first 4
contribute in the unitary case.}
\end{figure}
each diagram in the three tables.

\begin{center}
\begin{tabular*}{0.95\textwidth}{|p{0.942\textwidth}|}
\hline
\begin{eqnarray*}
3F_{a}&=&F_{d}=-2\sum_{q_1,q_2,q_3}\frac{2(\PP_1+\PP_2+\PP_3)+3i\omega}
{\PP_1\PP_2\PP_3}\\
F_{b}&=&\frac{1}{4}\sum_{q_1,q_2,q_3}\frac{(\PP_{12}+\PP_{23})^2}
{\PP_1\PP_2\PP_3\PP_{123}}\\
F_{c}&=&2\sum_{q_1,q_2,q_3}
\frac{(\PP_1+\PP_2+i\omega)(\PP_1+\PP_3+i\omega)}{\PP_1^2\PP_2\PP_3}\\
F_{e}&=&\frac{1}{2}\sum_{q_1,q_2,q_3}
\frac{(\PP_{12}+\PP_{23})(\PP_{13}+\PP_{32})}{\PP_1\PP_2\PP_3\PP_{123}}\\[-1cm]
\end{eqnarray*}
{\small\begin{tabbing}\hspace*{.8cm}{\bf Table 1. }\=
Three-loop order contributions to $\widetilde F(\omega)$:
an overall factor of $(\Delta/2\pi)^2$ should\\[-.3cm]\> be attached to each 
contribution.
\end{tabbing}}\\
\hline
\end{tabular*}
\vspace{1cm}

\begin{tabular*}{0.95\textwidth}{|p{0.942\textwidth}|}
\hline
\begin{eqnarray*}
p_a&=&\frac{1}{2}p_e=p_f=
-2\sum_{q_1,q_2,q_3}\frac{2(\PP_1+\PP_2+\PP_3)+3i\omega)}
{\PP_1^2\PP_2\PP_3}\\
p_b&=&\sum_{q_1,q_2,q_3}\frac{(\PP_{12}+\PP_{23})^2}
{\PP_1^2\PP_2\PP_3\PP_{123}}\\
p_c&=&4\sum_{q_1,q_2,q_3}\frac{(\PP_1+\PP_2+i\omega)(\PP_1+\PP_3+i\omega)}
{\PP_1^3\PP_2\PP_3}\\
p_d&=&4\sum_{q_1,q_2,q_3}\frac{(\PP_1+\PP_2+i\omega)(\PP_2+\PP_3+i\omega)}
{\PP_1^2\PP_2^2\PP_3}\\
p_g&=&2\sum_{q_1,q_2,q_3}\frac{(\PP_{12}+\PP_{23})(\PP_{13}+\PP_{32})}
{\PP_1^2\PP_2\PP_3\PP_{123}}\\
p_h&=&4\sum_{q_1,q_2,q_3}\frac{(\PP_1+\PP_2+i\omega)}{\PP_1^2\PP_2\PP_3}\\
p_i&=&-4\sum_{q_1,q_2,q_3}\frac{1}{\PP_1\PP_2\PP_3}\\
p_j&=&\sum_{q_1,q_2,q_3}\frac{(\PP_{12}+\PP_{23})}
{\PP_1\PP_2\PP_3\PP_{123}}\\[-1cm]
\end{eqnarray*}
{\small\begin{tabbing}\hspace*{.8cm}{\bf Table 2. }\=
 Three-loop order contributions to $\widetilde p(\omega)$:
an overall factor of $L^{-d}(\Delta/2\pi)^2$ \\[-.3cm]\> should
 be attached to each 
contribution.
\end{tabbing}}\\
\hline
\end{tabular*}
\end{center}

\begin{center}
\begin{tabular*}{0.95\textwidth}{|p{0.942\textwidth}|}
\hline
\begin{eqnarray*}
Q_a&=&2Q_b=Q_h=Q_i=Q_j=Q_k=2i\sum_{q_1,q_2,q_3}\frac{1}{\PP_1^2\PP_2\PP_3}
\\
Q_c&=&Q_l=i\sum_{q_1,q_2,q_3}\frac{1}{\PP_1\PP_2\PP_3\PP_{123}}\\
Q_d&=&Q_m=-2i\sum_{q_1,q_2,q_3}\frac{(\PP_{12}+\PP_{23})}
{\PP_1^2\PP_2\PP_3\PP_{123}}\\
Q_e&=&-4i\sum_{q_1,q_2,q_3}\frac{(\PP_1+\PP_2+i\omega)}
{\PP_1^3\PP_2\PP_3}\\
Q_f&=&-2i\sum_{q_1,q_2,q_3}\frac{(\PP_1+\PP_3+i\omega)}
{\PP_1^2\PP_2^2\PP_3}\\
Q_g&=&-2i\sum_{q_1,q_2,q_3}\frac{(\PP_1+\PP_2+i\omega)} 
{\PP_1^2\PP_2^2\PP_3}
\end{eqnarray*}
{\small\begin{tabbing} \hspace*{.8cm}{\bf Table 3. }\=
Three-loop order contributions to $\widetilde Q(\omega)$:
an overall factor of $(\Delta^2/2\pi L^d)$ \\[-.3cm]\> should
 be attached to each 
contribution.
\end{tabbing}}\\
\hline
\end{tabular*}
\end{center}

The labels in the tables correspond
to those in Figs.\ 8--10, and in the figure captions we describe which 
diagrams are made of diffusons only and thus contribute in the 
unitary ($\beta=1$) case. Obviously, all the diagrams contribute in the 
orthogonal ($\beta=2$) case. Results for the symplectic symmetry class
($\beta=4$) are practically the same as for the orthogonal case but we do
not list them here to avoid complications with the coefficients. 

The starting point for our diagrammatic analysis is Eq.\ (\ref{p=Q})
which we now rewrite as follows. The Fourier transform, 
Eq.\ (\ref{Q}),
  of the function $Q(t)$ is split into the reducible and irreducible
parts (Eqs.\ (\ref{Quc}) and  (\ref{Qc}), respectively). As only 
$\widetilde Q(\omega)$, Eq.\ (\ref{tilde}), contributes to the irreducible
part, we obtain 
\begin{eqnarray}
\label{exact2}
\frac{\partial}{\partial (i\omega)}\left[\frac{1}{L^d}
\frac{\partial}{\partial (i\omega)}
F(\omega)- \widetilde{p}(\omega)\right]
=\frac{i}{\pi}\widetilde Q(\omega)
\end{eqnarray}
This is the exact relation which holds to all orders
in perturbation theory. It allows us to check the accuracy of our 
diagrammatics up to three loop order for both the 
unitary and orthogonal cases. We do this   by first substituting 
the two-loop results of Eq.\ (\ref{twoloop}) - which is quite straightforward,
and then the three-loop data from the tables which requires some significant
algebra.  We verify that this identity holds with our diagrammatic
results which gives us confidence in their accuracy. 

The next step is to see whether our projected $\omega$-space factorization
occurs to two and three loop order in both the orthogonal and unitary cases.
To check  Eq.\ (\ref{twofac}) up to  two-loop order, we calculate both factors
in the r.h.s.\ to the first order only. This is simple and yields
\begin{eqnarray*}
\widetilde Q_2(\omega)=-i\left(\frac{2}{\beta}\right)^2
\sum_{q_1,q_2}\frac{1}{\bigl(\PP_1\bigr)^2\,\PP_2}
=-\frac{i(2\pi)^2}{\Delta}\, \widetilde{p}_1(\omega)\widetilde{R}_1(\omega)\,,
\end{eqnarray*}
so that the factorization (\ref{twofac}) is exact in this order. 
The reason why this is so simple is that with only two momenta $q_1$
and $q_2$ there is no way for the momenta to become ``entangled'', and
so there is only really one possible functional form. The fact that the
numerical coefficients match up exactly is the important thing. The 
three loop case is more involved because now the entanglement can 
occur, and this leads to the factorization not being exact. For both
the unitary and orthogonal case, however, we get the same functional
form in the remainder,
\begin{eqnarray*}
& &Q_3(\omega)+\frac{i(2\pi)^2}{\Delta}\left[
\widetilde{R}_1(\omega)\widetilde{p}_2(\omega)+
\widetilde{R}_2(\omega)\widetilde{p}_1(\omega)\right]\\
&=&-\frac{2i\Delta^2}{\pi\beta^2L^d} 
\sum_{q_1,q_2,q_3}\left\{
\frac{4}{
\bigl(\PP_1\bigr)^2\,\PP_2 \,\PP_3}
-\frac{2\bigl(\PP_{12}+\PP_{23}\bigr)}
{\bigl(\PP_1\bigr)^2\,\PP_2 \,\PP_3\,\PP_{123}}+
\frac{1}{\PP_1\,\PP_2 \,\PP_3\,\PP_{123}}
\right\}\,.
\end{eqnarray*}
The remainder can then be algebraically manipulated to give
\begin{eqnarray}
\label{threefac1}
\frac{2\Delta^2}{3\pi\beta^2L^d} \,\frac{\partial}
{\partial(i\omega)}
\sum_{q_1,q_2,q_3}\left\{
\frac{i\omega}{\PP_1\,\PP_2 \,\PP_3\,\PP_{123}} \right\}\,.
\end{eqnarray}
At the three loop level we see that the factorization is not exact,
but that the remainder term is simple and its momentum
integrals are fully convergent. Certainly for the orthogonal case
many terms have been removed to yield this remainder. We next note that
in the special case of two dimensions this remainder is zero, and the
factorization is exact. This is because for $d=2$ dimensional analysis
shows that the term inside the derivative is a constant -- it is of
the form $(-i\omega)^0$, and no logarithmic singularities are present -- 
and so one gets zero upon taking derivative. Even if the remainder were
not able to be written as a derivative, but just as a sum of terms with
no logarithmic singularities, it would yield zero. This is because
dimensional analysis shows that the result is of the form $a/(-i\omega)$,
where $a$ is a real constant. Since we have to take the real part of
this to get $R(\omega)$ this would give zero. However in this case there
would be a constant contribution to the Fourier transform $K(t)$,
because $a/(-i\omega)$ does actually have a real part proportional to
$\delta(\omega)$ -- this is exactly what happens in the case of the 1-loop
contribution in $d=2$. We note that our factorization is exact to 3-loop
order in $d=2$ even in the sense of getting the constant term in $K(t)$
correct. 
 
It seems to us that the
exactness of factorization up to three loop order in $d=2$ for both
orthogonal and unitary cases is no accident, and we conjecture that
this result persists to all orders in perturbation theory. Obviously
such a conjecture cannot be proved using order-by-order analysis
(although, of course, it could be disproved this way), so any attempt
to verify this will require analysis of the structure of $R(\omega)$, 
$p(\omega)$, and $Q(\omega)$ diagrams.

\subsection{Comparison of the Two Factorization Schemes}

In this section we will compare the two factorization schemes that
we have introduced in this paper: the $t$-space scheme that arose from
consideration of the Brownian motion model of section (III), 
and the $\omega$-space
scheme that arose in the diagrammatic analysis of section (VI).
We have shown that the $\omega$-space scheme is exact to 2-loop order
in all dimensions, and to 3-loop order in the 2d case. We will now
examine the validity of the $t$-space scheme. This involves very little
extra work because most of the algebraic manipulation has already
been performed in the $\omega$-space analysis. We first compare the two 
relations by writing both of them in $t$-space to yield
\begin{mathletters}
\label{compare}
\begin{eqnarray}
\label{espace}
K(t)-(2\pi\hbar\rho)^{-1}tp(t) &=& -(\pi\hbar\rho)^{-1}
\int_{0+}^t dt' K(t')p(t-t') \\
\label{tspace}
K(t)-(2\pi\hbar\rho)^{-1}tp(t) &=& -(\pi\hbar\rho)^{-1}K(t)\int_0^t p(t')
\end{eqnarray}
\end{mathletters}
where, of course, we know the region of validity of the first formula.
To investigate the $t$-space factorization we need only compare the
r.h.s. of the above equations. In the 2-loop case the $K(t)$ and $p(t)$
in the r.h.s. will both be of 1-loop order, and we know that
$p_1(t)\propto t^{-d/2}$ and $K_1(t)\propto t^{1-d/2}$.
We find that the two equations above only agree for $d=2$,
where $K_1(t)=1$. The $t$-space factorization is therefore exact
to 2-loops only in 2d, and we restrict ourselves to the 2d case from
now on.

To look at the $t$-space scheme to 3-loop order in 2d we note that
we can have the combinations $K_1$, $p_2$, and $K_2$, $p_1$. For the
unitary case $K_2(t)=0$, so this leaves only the first contribution,
and since $K_1(t)=1$ the two factorizations become the same. For the
orthogonal case we have to look at the second contribution. We find
that the two equations differ by a constant. For the purpose
of calculating the asymptotic behaviour of $R(\omega)$, constant
terms in $K(t)$ have no effect, so that the $t$-space
factorization works up to 3-loop order in 2d for both orthogonal and
unitary cases.

At this point it seems that the $\omega$-space scheme may be 
perturbatively slightly more
accurate than the $t$-space scheme in that it is correct in 2-loops
for all dimensions, and at 3-loop order it gets the constant term in
$K(t)$ correct in the orthogonal case. However we are still justified
in saying that both schemes are correct to 3-loop order in 2d.

\section{Summary}

In this paper we have examined spectral correlations in 
disordered conductors,
starting from the idea (Eq.\ \ref{lambda}) that two samples 
with impurity configurations differing by an infinitesimal 
amount should be statistically equivalent. In the first instance, 
this equivalence yields an identity relating two-point to three-point 
correlation functions; it is useful only if one can decouple the 
latter. We argue that a decoupling based on the Kirkwood 
superposition approximation is physically reasonable provided 
one is interested only in correlations at scales large compared 
to the mean level spacing. 
Within the approximation, we
express (Eq.\ \ref{K1}) both the non-parametric and the parametric spectral
form factor in terms of the quantum return probability
for a spreading wavepacket.
We test the decoupling by
calculating corrections, using the standard diagrammatic perturbation
theory for disordered conductors to expand about the
metallic limit in inverse powers of the dimensionless conductance, $g$.
We show that in two-dimensional systems, the case of greatest interest,
there are no corrections to order $g^{-3}$. We believe that
the results we obtain from this approach should be useful rather
generally, and especially when a diagrammatic analysis is not
straightforward, as at the Anderson transition; the implications of our
work in that regime will be discussed elsewhere.

\acknowledgements
We thank V.~E.~Kravtsov and B.~D.~Simons for numerous
useful discussions. Support by the EPSRC under grants Nos. GR/GO 2727 (J.T.C.)
and GR/J35238 (I.V.L.\ and R.A.S.) is gratefully acknowledged.
I.V.L.\ acknowledges the hospitality of the ITP in Santa Barbara where part of
this work was performed, and partial support by the NSF under grant 
No. PHY94-07194.


\begin{references}
\bibitem{bmcom2}
In the ergodic regime, this expression gives only the envelope of the TLCF as
  the oscillations in $R(s)$ \cite{RMT} are due to singularities in $K(t)$ near
  $t_{\!_H}$, which cannot be accounted for semi-classically.
\end{references}
\end{document}